\documentclass[fleqn,usenatbib]{mnras}

\usepackage[T1]{fontenc}
\usepackage{ae,aecompl}

\usepackage{graphicx}	
\usepackage{amsmath}	
\usepackage{amssymb}	
\usepackage[utf8]{inputenc}
\usepackage{xcolor}
\usepackage{newtxtext,newtxmath}

\newcommand{\mygal}
 {the dIGal}
\newcommand{\mygalsimple}
 {dIGal}
\newcommand{\RN}[1]{%
  \textup{\uppercase\expandafter{\romannumeral#1}}%
}

\title[The metal-poor dwarf galaxy next to Mrk 1172]{The metal-poor dwarf irregular galaxy candidate next to Mrk~1172}

\author[Lassen et al.]{Augusto E. Lassen$^{1}$\thanks{augusto.lassen@ufrgs.br}, Rogerio Riffel$^{1}$, Ana L. Chies-Santos$^{1}$, Evelyn Johnston$^{2,3}$,
\newauthor 
Boris Häußler$^{4}$, Gabriel M. Azevedo$^{1}$, Daniel Ruschel-Dutra$^{5}$, Rogemar A. Riffel$^{6}$
\\
$^{1}$Universidade Federal do Rio Grande do Sul, Departamento de Astronomia, Av. Bento Gonçalves 9500, Porto Alegre, RS, Brazil\\
$^{2}$Universidad Diego Portales, N\'ucleo de Astronom\'ia de la Facultad de Ingenier\'ia y Ciencias, Av. Ej\'ercito Libertador 441, Santiago, Chile\\
$^{3}$Pontificia Universidad Católica de Chile, Institute of Astrophysics, Av. Vicuña Mackenna 4860, 7820436 Macul, Santiago, Chile\\
$^{4}$European Southern Observatory, Alonso de Córdova 3107, Vitacura, Santiago, Chile\\
$^{5}$Universidade Federal de Santa Catarina, Departamento de Física, \textit{P.O. Box} 476, 88040-900, Florianópolis, SC, Brazil\\
$^{6}$Universidade Federal de Santa Maria, Departamento de Física, Centro de Ciências Naturais e Exatas, 97106--900, Santa Maria, RS, Brazil}
\date{Accepted XXX. Received YYY; in original form ZZZ}

\pubyear{2021}
\begin{document}

\label{firstpage}
\pagerange{\pageref{firstpage}--\pageref{lastpage}}
\maketitle

\begin{abstract}
In this work we characterise the properties of the object SDSS~J020536.84-081424.7, an extended  nebular region with projected extension of 14\,$\times$\,14\, kpc$\rm ^2$ in the line of sight of the ETG Mrk~1172, using unprecedented spectroscopic data from MUSE. 
We perform a spatially resolved stellar population synthesis and estimate the stellar mass for both Mrk~1172 (1 $\times 10^{11}$ M$_{\odot}$) and our object of study ($3 \times 10^{9}$\,$M_{\odot}$). While the stellar content of Mrk~1172 is dominated by an old ($\sim$ 10\,Gyr) stellar population, the extended nebular emission has its light dominated by young to intermediate age populations (from $\sim$ 100\,Myr to $\sim$ 1\,Gyr) and presents strong emission lines such as: H$\beta$, [\ion{O}{iii}] $\lambda,\lambda$4959, 5007\AA, H$\alpha$, [\ion{N}{ii}] $\lambda,\lambda$6549, 6585\AA\ and [\ion{S}{ii}] $\lambda,\lambda$6717, 6732\AA.  Using these emission lines we find that it is metal-poor (with $Z \sim$ 1/3\,$Z_{\odot}$, comparable to the LMC) and is actively forming stars ($0.70 \ M_{\odot}$\,yr$^{-1}$), especially in a few bright clumpy knots that are readily visible in H$\alpha$. The object has an ionised gas mass $\geq 3.8 \times 10^{5}$\,$M_{\odot}$. Moreover, the motion of the gas is well described by a gas in circular orbit in the plane of a disk and is being affected by interaction wtih Mrk~1172. We conclude that SDSS~J020536.84-081424.7 is most likely a dwarf irregular galaxy (\mygal).
\end{abstract}

\begin{keywords}
galaxies: dwarf -- \ion{H}{ii} regions -- ISM: abundances
\end{keywords}

\section{Introduction}

Dwarf galaxies are the most abundant class of galaxies in the Universe, and play a fundamental role in models of galaxy formation and evolution. In the hierarchical framework of galaxy formation, they are the building blocks of larger objects, contributing to the assembly of larger galaxies via successive mergers \citep[e.g.][and references therein]{mateo1998,revaz2018,digby2019}. Since these galaxies are numerous, they probe many different environmental conditions and are sensitive to perturbations such as stellar feedback, for example, because their shallower gravitational potential wells make the pressure of the gas within these galaxies lower when compared to more massive galaxies \citep{navarro1996,mashchenko2008}. While some dwarf galaxies may evolve in isolation, many of them are found within systems where the effects of tidal and ram-pressure forces, for example, leave imprints in their Star Formation Histories (SFHs). In the former case, these objects represent ideal laboratories to study internal drivers of galaxy evolution, like gas accretion \citep{vanzee2001,vanzee2006,bernard2010,gonzalez2014}. In the latter case, these compose an important set of galaxies to study the effects of environment in the determination of the mass and the structure of satellites \citep{mayer2001,kazantzidis2010,fattahi2018,Steyrleithner20}.

Traditionally, dwarf galaxies are classified in two main categories, the dwarf Irregular (dI) and the dwarf Spheroidal (dSph). The dI galaxies are rich in gas and actively forming stars, usually located in the field, while dSph galaxies do not present significant star formation activity \citep{mayer2001,gallart2015}. The differences between both categories of dwarfs are usually based on current properties, and probably do not reflect
their evolutionary histories. Since dIs and dSphs share many structural and evolutionary properties, the categorisation is most likely to be related to the fact that in dSphs the Star Formation (SF) has ceased recently, while in dIs this SF persists until the present day \citep{skillman1995,kormendy2011,kirby2013}.

In general, dIs are metal-poor, and the abundance of heavy elements in these galaxies, measured from their \ion{H}{ii} regions, lies in the range of 1/3 -- 1/40\,$Z_{\odot}$ \citep{kunth2000}. As examples of metal-poor dIs we have the SMC and the LMC, with a metallicity of roughly 1/8\,Z$_{\odot}$ and 1/3\,Z$_{\odot}$, respectively \citep{kunth2000}. There are also the Blue Compact Dwarf Galaxies (BCGDs), which can reach lower metallicities in their ISMs, as it is the case of I~Zw~18. This BCDG has the lowest nebular Oxygen abundance among star forming galaxies in the nearby Universe, with Z$\sim$\,1/50\,Z$_{\odot}$ \citep{aloisi2007}.

Understanding the chemical evolution of dwarf galaxies is essential to constrain models of galaxy formation and evolution, since the fraction of heavy elements within a galaxy is not only related to secular processes, but can also carry imprints of past merging episodes \citep{lequeux1979,skillman1989,croxall2009}. The mass-metallicity relation is a widely studied empirical trend that holds from dwarfs to very massive galaxies, where the most massive galaxies are also more metal-rich \citep{rubin1984,pilyugin2000,tremonti2004,andrews2013}. An explanation for this relation is the easier retention of the metallic content by galaxies with deeper gravitational potential wells, since the mechanisms of gas transport such as gas accretion (infall) and winds (outflows), which are able to affect the metallicity of a galaxy, depend on the mass of the system \citep{gibson1997,dalcanton2007}. Thus, dwarf galaxies are excellent laboratories to study the chemical evolution of metal deficient galaxies and offer unique conditions to improve our understanding of galaxy formation and evolution.

In this paper we analyse an intriguing object in the vicinity of the massive Early-Type Galaxy (ETG) Mrk~1172. This object has available photometric data in the optical (Sloan Digital Sky Survey, SDSS) and in the NUV and FUV (Galaxy Evolution Explorer, GALEX), but, to the best of our knowledge, has no previous detailed analysis on the literature (see \S~\ref{oursource}). In Section~\ref{sec:2} we introduce the data and the adopted analysis methodologies, in Section~\ref{sec:3} we briefly describe the techniques and present the results obtained, in Section~\ref{sec:4} we discuss our results and the conclusions are presented in Section~\ref{sec:5}. Throughout this work, we adopt
$H_{0} = 69.32 \ \mathrm{km} \ \mathrm{s}^{-1} \ \mathrm{Mpc}^{-1}$, $\Omega_{m} = 0.2865$, $\Omega_{\Lambda} = 0.7135$ \citep{hinshaw2013} and Oxygen abundance as a tracer of the overall gas phase metallicity, using the two terms interchangeably, and a solar Oxygen abundance of $\mathrm{log} (O/H) + 12 = 8.76$ \citep{steffen2015}.

\section{Observations and data reduction}\label{sec:2}
In this work we present integral field spectroscopy (IFS) of the Mrk~1172 region (J020536.18-081443.23) from Program-ID 099.B-0411(A) (PI: Johnston). The data were obtained using the Multi-Unit Spectroscopic Explorer \citep[MUSE,][]{bacon2010} on the Very Large Telescope (VLT) in the wide-field mode (WFM), covering the nominal wavelength range of 4650 -- 9300 \AA \ with mean spectral sampling of 1.25 \AA, a FoV of 1x1~arcmin$^{2}$, angular sampling of 0.2 arcsec and seeing of $\sim$ 1.4 arcsec. Mrk~1172 was observed on the nights of 2018 August 10th and 2018 October 1st, with a total exposure time of 1.6\,hours split over 6 exposures. The images were rotated and offset to remove the effect of slicers and channels.

\subsection{Data Reduction}
For each night of observations a standard star was observed for flux and telluric calibrations, sky flats were taken within a week of the observations and an internal lamp flat was taken immediately before or after each set of observations. This lamp flat image was used to correct for the time and temperature dependent variations in the background flux level of each CCD. Additional bias, flat field and arc images were observed the morning after each set of observations.

The data were reduced using the ESO MUSE pipeline \citep{weilbacher2020} within the ESO Recipe Execution Tool ({\sc EsoRex}) environment \citep{ESOteam2015}. First we created master bias and flat field images and a wavelength solution for each detector and for each night of observations. The flux calibration solution obtained from the standard star observations and the sky flats were then applied to the science frames as part of the post-processing steps. The reduced pixel tables created for each exposure by the post-processing steps were combined to produce final data cube. Since the sky-subtraction applied as part of the EsoRex pipeline leaves behind significant residuals that may contaminate the spectra of faint sources, particularly at the NIR wavelengths,  we applied the Zurich Atmosphere Purge \citep[ZAP,][]{zap2016} to the final data cube in order to improve sky subtraction and minimize these artefacts.

\subsection{Data specifications}\label{oursource}
\begin{figure*}
	\includegraphics[scale=0.73]{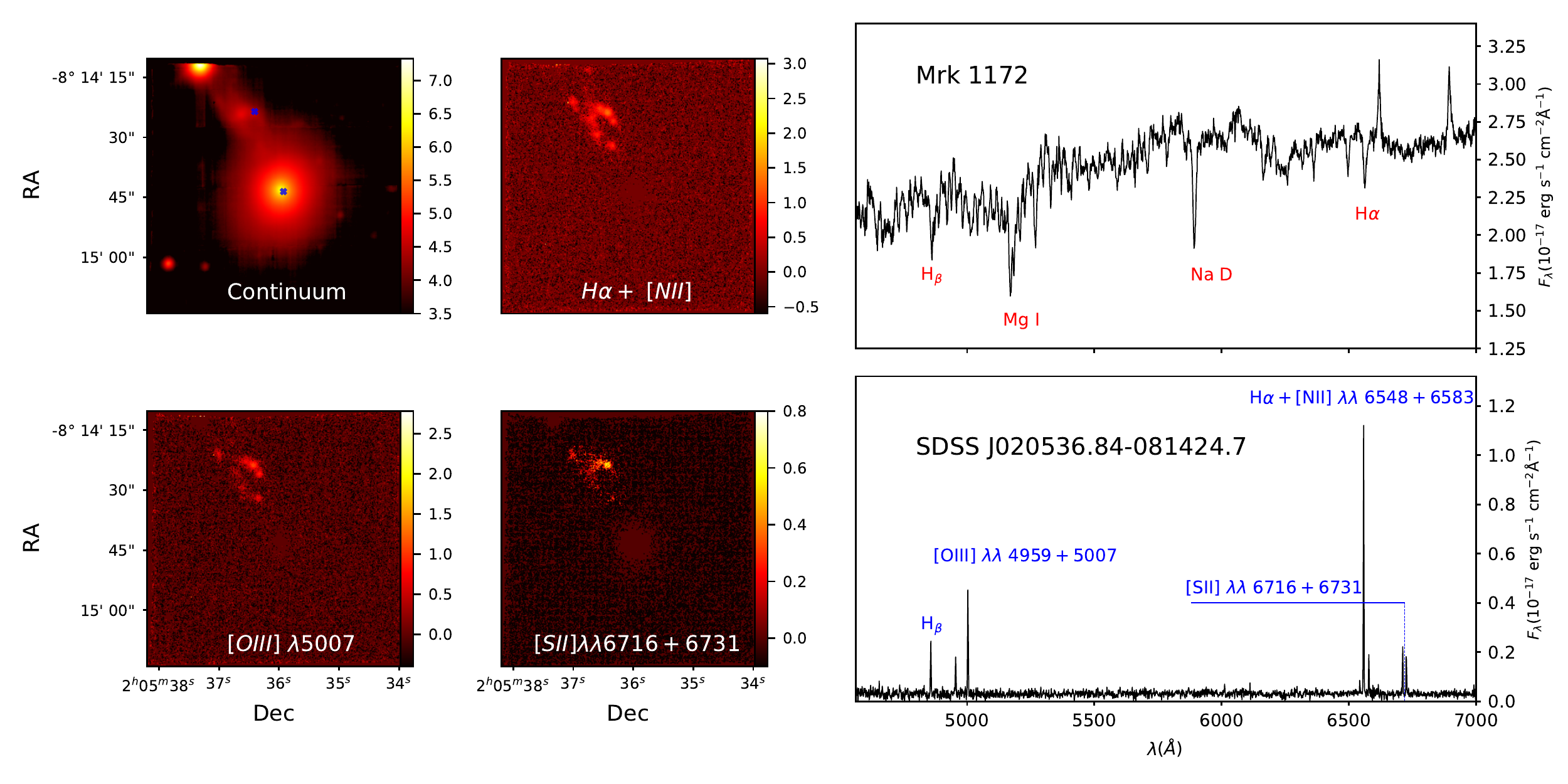}
    \caption{The MUSE FoV continuum centered in Mrk~1172 (top left), H$\alpha$+[\ion{N}{ii}]$\lambda \lambda$6550+6585 (top central), [\ion{O}{iii}]$\lambda 5007$ (bottom left) and [\ion{S}{ii}]$\lambda \lambda$6716+6731 (bottom central) wavelength ranges. The wavelength windows used to subtract the continuum from the flux of each emission lines are presented in section 2. The spectrum of Mrk~1172 highest-SNR spaxel is presented in top right panel, while in bottom right panel we present the spectrum for the highest-SNR spaxel for \mygal. The location of these spaxels within the FoV is indicated in blue in the top left panel. The colorbars represent in logarithmic scale the flux in each image, in units of $10^{-20}$\,erg\,s$^{-1}$\,cm$^{-2}$. 
    }
    \label{fig:Figura1}
\end{figure*}

During the inspection the data cubes from the MUSE Program-ID 099.B-0411(A), we identified an extended nebular region in the FoV of the ETG Mrk~1172 ($\alpha$= 02$^{\mathrm{h}}$\,05$^{\mathrm{m}}$\,36.19$^{\mathrm{s}}$, $\delta$=\,-08$^{\mathrm{h}}$\,14$^{\mathrm{m}}$\,43.25$^{\mathrm{s}}$). This region occupies a projected area of approximately 92\,arcsec$^{2}$ in the observed field of view (FoV), calculated by defining a rectangle containing the correspondent object. The FoV is shown in the top left panel of Fig.~\ref{fig:Figura1}. We inspected this FoV in SDSS, and found that the bright source to the north-east of MRK~1172 (SDSS~J020537.54-081411.5) and the two sources to the south (SDSS~J020538.07-081501.2 and SDSS J020537.43-081502.2) are stars. The SDSS data also showed many small galaxies around Mrk~1172, most of them not visible in our data, indicating that this system could perhaps be a fossil group. Our object of interest is listed as SDSS~J020536.84-081424.7, and has photometric information, though no spectra is available. This object also appears in the GALEX catalogue as GALEX J020536.7-081424, where photometric information on the NUV and FUV is available. For simplification, throughout this paper we will refer to it as the Dwarf Irregular Galaxy (\mygalsimple).

To illustrate the spectra of \mygal\ and Mrk~1172, we selected the spaxels with highest SNR in each galaxy, the locations of which are represented by blue crosses in the top left panel of Fig.~\ref{fig:Figura1}. Both spectra are shown in the right panel of Fig.~\ref{fig:Figura1}, with main emission and absorption lines labelled. There are two prominent lines in Mrk~1172 spectrum red-wards of the H$\alpha$  absorption line that resemble emission lines. Comparing our spectrum with the one available in SDSS, we notice that these features are absent in the SDSS spectrum. These features are not normal sky lines since they are broad and do not appear in the spaxels corresponding to sky background. There are other lines that appear in the red part of the spectrum, but curiously they are present only on the data corresponding to the second night of observation. We believe that these features are actually artefacts from the telluric correction. Usually telluric lines appear as broad absorption lines. However, in the case observations from the second night, the telluric lines are stronger in the standard star datacube than in the science datacube, causing an overcorrection of these features, and thus appearing as emission lines in the final datacube. 

In the left of panel of Fig.~\ref{fig:Figura1} we present the FoV now centred on the wavelength range corresponding to the main emission lines seen in \mygal\ spectrum, subtracting the adjacent continuum. The emission lines and the continuum regions used in Fig.~\ref{fig:Figura1} are listed below:

\begin{itemize}
\item $ \ $ H${\alpha}$ + [\ion{N}{ii}]: 6525--6535 \AA, 6590--6600 \AA
\item $ \ $ [\ion{O}{iii}] $\lambda$5007: 4990--5000 \AA, 5015--5044 \AA
\item $ \ $ [\ion{S}{ii}] : 6650-6700 \AA, 6800--6900 \AA
\end{itemize}

Using the strong emission lines in the spectra of the \mygal\ we were able to measure its redshift as $z=0.04025\pm0.00003$. In SDSS, Mrk~1172 has a reported redshift of $z=0.04115$ \citep{SDSS2012}. The redshift measured for Mrk~1172 with MUSE observations is consistent with this previous measurement, and this is the value used throughout the paper.
\section{Analysis}\label{sec:3}
\subsection{Stellar population fitting}
A spatially resolved stellar population synthesis analysis is essential to reveal information regarding the formation, evolution and current state of the observed systems. With such techniques we can obtain the star-formation history (SFH) of both Mrk~1172 and \mygal, as well as separating the stellar continuum/absorption features and the emisison lines from the gas. This analysis will allow us to estimate several properties of the galaxies, such as extinction, star formation rate (SFR) and other properties that shall be discussed further, and which can be illustrated via 2D maps \citep{cid2013,nicolas2018,nascimento2019}. 

To perform the stellar population synthesis we use the {\sc megacube} module, which was developed to work as a front-end for the {\sc starlight} code, operating in three main modules \citep{cid2005,nicolas2018,riffel2021}. Since {\sc starlight} is designed to operate with ASCII-format files, the spectrum of each spaxel needs to be extracted from the original \textit{fits} files, applying several pre-processing corrections, i.e., rest-frame spectrum shifting and galactic extinction correction. We used the dust maps from \citet{schlegel1998} and the CCM reddening extinction law \citep[using $R_{V} = 3.1$,][]{cardelli1989,odonnell1994}. 

In order to increase the SNR of the individual spaxels, we have binned the original data cube 2x2 along the spatial direction. In order to increase the SNR of the individual spaxels, we have binned the original data cube 2x2 along the spatial direction. Since the spatial coverage of MUSE is large, many spaxels within the FoV cover empty regions in the sky or correspond to objects in the FoV that we would like to mask out from the analysis, such as stars. Therefore, it is useful to create a $2D$ boolean mask to flag valid and invalid spaxels. However, to create the mask we must adopt a reliable criteria to separate valid spaxels from the invalid ones. In this work we adopt the following criteria:

\begin{itemize}

\item The flux vector of the individual spaxel must present 5~\% or lower of negative and infinite numbers within the array in order to avoid the noisy spaxels, especially from the edges. In the case of spaxels considered to be valid by the above criterion but that still have few invalid numbers, we replace this numbers by applying an interpolation with neighbouring valid values.

\item The maximum of the flux (emission or absorption lines) must be at least 1.5 times greater than the standard deviation for that flux vector. In this way only spaxels with high SNR and/or strong emission and absorption lines are set as valid.
\end{itemize}

\begin{figure}
	\includegraphics[width=\columnwidth]{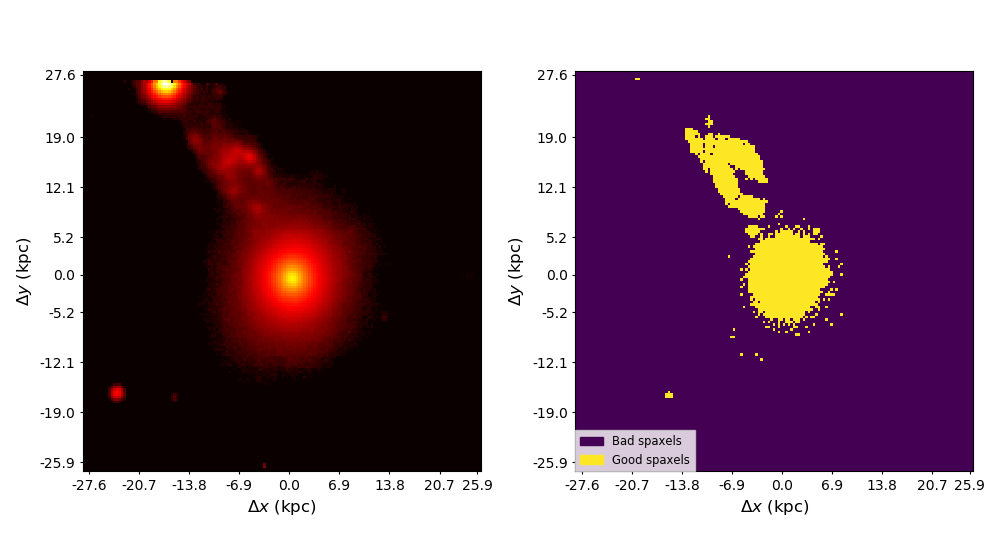}
    \caption{Comparison between Mrk 1172 FoV in continuum plus H${\alpha}$ emission (left panel) with a binary map (left panel), where the yellow regions represent the spaxels used in the synthesis, while the purple spaxels are discarded. Undesired high SNR spaxels, such as those corresponding to the stars in the FoV were masked manually. The coordinates presented represent the offset in kpc from the reference spaxel of the FoV, centered in Mrk~1172.}
    \label{fig:2d_mask}
\end{figure}

\begin{figure*}
    \centering
    \includegraphics[scale=0.75]{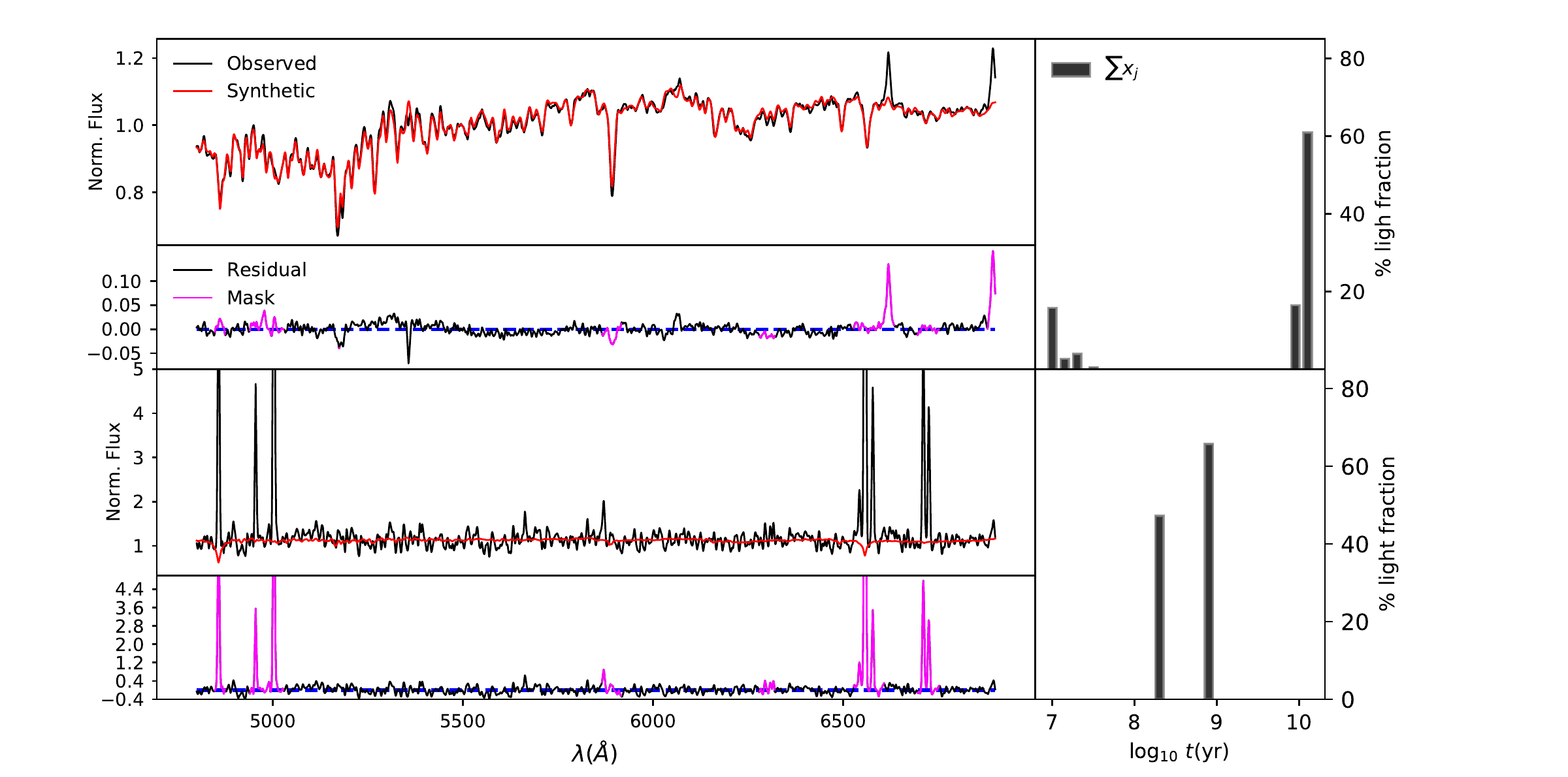}
    \caption{Synthesis results for the highest-SNR spaxel for Mrk~1172 and \mygal. \textit{Top left panel}: In black, the observed spectrum of Mrk~1172, where the flux is normalised using $F(\lambda_{0})$. In red, the synthetic spectrum (the best-fit from the stellar population synthesis). \textit{Second top left panel:} Residual spectrum for Mrk~1172 in black, with the regions masked by sigma clipping method in magenta. At the bottom left panels, the same for \mygal. \textit{Top right panel}: Histogram presenting the luminosity-weighted contribution of each stellar population for Mrk~1172 best fit and their ages. At the bottom right panels, the same for \mygal.}
    \label{fig:synth_results}
\end{figure*}

The values used to create the criteria were obtained after several tests, for example by comparing the final 2D mask with the continuum image added with $\mathrm{H}\alpha$ emission, as shown in Fig.~\ref{fig:2d_mask}. Finally, each spaxel was convoluted with a Gaussian function\footnote{The new Gaussian kernel was obtained by subtracting the data and model $\sigma$ values in quadrature and taking its root square mean value.} in order to match the spectral resolution of the target spectra (R$\sim$ 2850 at 7000\AA) with that of the Simple Stellar Population (SSP) models \citep[which have a FWHM = 2.51\AA\ resolution][]{vazdekis2010,vazdekis16} and rebinned them to $\delta\lambda$ = 1\,\AA.
 For the spaxel meeting the above criteria and with the corrections applied we performed the stellar population synthesis. We used the Granada-Miles SSP models computed with the PADOVA200 isochrones and Salpeter initial mass function \citep{vazdekis2010,cid2014}. We adopted 21 ages (0.001, 0.0056, 0.01, 0.014, 0.020, 0.031, 0.056, 0.1, 0.2, 0.316, 0.398, 0.501, 0.638, 0.708, 0.794, 0.891, 1, 2, 5, 8.9 and 12.6 Gyr) and 4 metallicities (0.19, 0.39, 1.0 and 1.7 $Z_{\odot}$). The fit was performed in the 4800--6900\,\AA\ spectral range with the normalisation point at $\lambda_0$ = 5600\,\AA, a spectral region free of absorption/emission lines. The prominent lines present in Mrk~1172 spectra do not interfere in the stellar population synthesis, since they were masked out. For safety we also used the the sigma clipping algorithm of {\sc starlight}, in order to remove any additional spurious features (see Fig.~\ref{fig:synth_results})

With the stellar population synthesis done, we inspect the result for individual spaxels in Mrk~1172 and \mygal. The locations of the spaxels with the highest SNR (approximately 90 and 7, respectively) spaxels in each target are marked in blue in Fig.~\ref{fig:Figura1}. In Fig.~\ref{fig:synth_results} we show examples fits to these spectra, with Mrk~1172 in the top panels and for  \mygal\ in the bottom panels. We present the observed spectrum in black with the best-fit synthetic spectrum in red, and below we show the residual spectrum and the regions masked by the sigma clipping method from {\sc starlight} for both cases. In the right panels we show the histograms of the contribution weighted by luminosity of each stellar population with its respective ages.
Mrk~1172 has dominant old stellar populations ($\sim 10^{10}$\,yrs), with the contribution of an artificial young stellar population ($\sim 10^{7}$\,yrs) that appears due to the AGN in this galaxy \citep{cid2005,riffel2009}. Meanwhile, \mygal\ is predominantly young, presenting two dominant stellar populations with ages between $\sim$100\,Myr and $\sim$1\,Gyr.

\begin{figure}
	\includegraphics[width=\columnwidth]{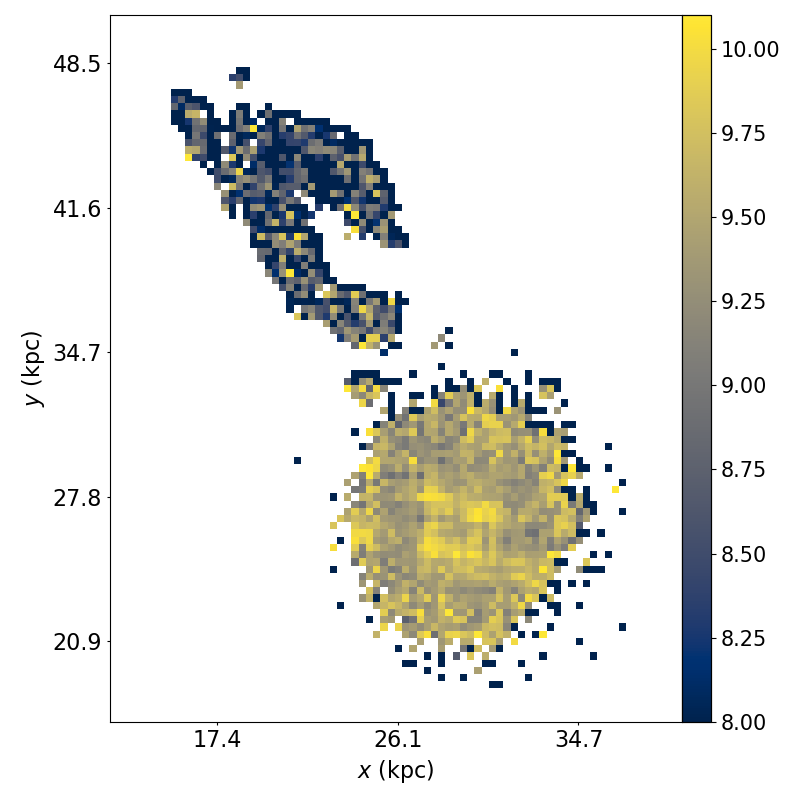}
    \caption{Mean luminosity-weighted age map for Mrk 1172 and \mygal\ in Gyr, obtained via spatially resolved stellar population synthesis.}
    \label{fig:mean_age}
\end{figure}

The luminosity-weighted mean age ($<t_{L}>$) for each spaxel can be obtained using (\citealt{cid2005}):

\begin{equation}
    <t_{L}> = \sum\limits_{j} a_{j} \ \mathrm{log} \ t_{j},
    \label{eq:age_L}
\end{equation}
where $t_{j}$ is the age of the j-th SSP model spectrum, and $a_{j}$ is a normalisation factor that takes into account the fact that the sum of the SSPs used to reproduce the input spectrum is not necessarily 100 \% \citep{cid2005}. The resulting map is presented in Fig.~\ref{fig:mean_age}, showing that \mygal\ is dominated by young and intermediate stellar populations, while Mrk 1172 is dominated by old stellar populations. This figure shows the spatial distribution of the information given by the histogram in Fig.~\ref{fig:synth_results}. It can be seen in Fig.~\ref{fig:mean_age}  that the stellar populations dominating the light emitted from  \mygal\ are considerably younger than those in Mrk~1172.

\subsection{Fitting of the emission-line profiles}
Having fitted the underlying spectrum of each spaxel, we subtract it from our observed data, resulting in pure emission line spectra. No emission lines were detected in locations corresponding to Mrk~1172, while strong emission features were found in the spectra of the \mygal, as seen in Fig.~\ref{fig:Figura1}. We measured the fluxes using the {\sc ifscube} \citep{ifscube} tool, which fits the emission-line profiles at each position by single Gaussian curves.

\subsubsection{Single gaussian fitting}
Fig.~\ref{fig:synth_results} shows that the synthetic spectrum produced through the stellar population synthesis models the continuum very well, thus we have not added a polynomial function to fit the continuum. The fitting of the emission lines was performed on the residual cube, obtained by the subtraction of the modelled continuum/absorption spectra from the cube used in the stellar population synthesis (i.e., a rest-frame spectrum, resampled in $\Delta \lambda = 1$ \AA\ and corrected for Galactic reddening). As a first approach, we assume that emission lines corresponding to transitions on the same atom (i.e., [\ion{O}{iii}] $\lambda$4959 and [\ion{O}{iii}] $\lambda$5007) belong to the same kinematic group\footnote{Many spectral features are physically linked, being produced in the same region of the target of study. Therefore, it is reasonable to assume that different transitions of the same atom were produced in the same region, and their emission lines share properties like velocity and velocity dispersion. When both lines share these properties, we say that they belong to the same kinematic group.}. Since the strongest emission lines in the spectrum of \mygal\ are from transitions of H, O, N and S, we use four different kinematic groups, one for each element.
We also used specific constraints for the line ratios, e.g. [\ion{N}{ii}] $\lambda$6583 = 3.06 $\times$ [\ion{N}{ii}] $\lambda$6548 and 0.4 < [\ion{S}{ii}]$\lambda$ 6717/[\ion{S}{ii}]$\lambda$ 6731 < 1.4 \citep{osterbrock2006}. The ratio of [\ion{O}{iii}] emission lines was kept free of constraints during the fitting process.

\begin{figure}
	\includegraphics[width=\columnwidth]{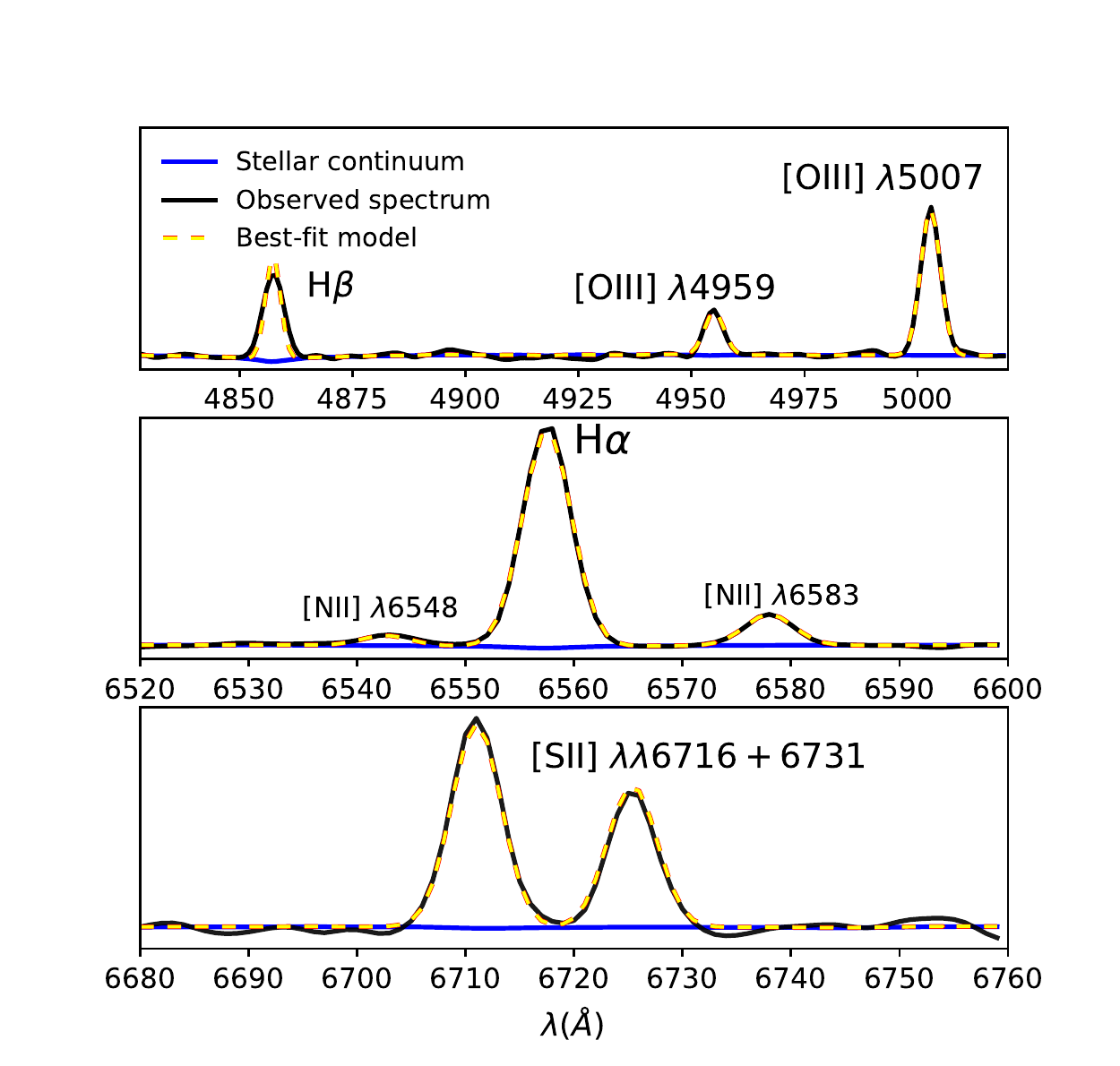}
    \caption{Emission lines in \mygal\ spectrum fitted using gaussian profiles with {\sc ifscube}. In black, the observed spectrum, in blue the continuum, and the dashed yellow line represents the model fit.}
    \label{fig:ifscube_fit}
\end{figure}

Fig.~\ref{fig:ifscube_fit} shows an example of the outcomes of this fitting procedure, where we present the spectral regions containing the fitted emission lines. The observed spectrum is shown as a solid black line, the stellar continuum is shown in the solid blue lines and the best-fit model is represented by a dashed yellow line. Besides fluxes, {\sc ifscube} also provides the gas kinematic parameters (gas velocity and velocity dispersion). In the next sections we analyse these quantities.

An important side note is that the S doublet, which can be used to estimate the electron gas density \citep[$n_e$,][]{osterbrock2006,ryden2015}, presents an emission line ratio that falls in the limit of sensibility of the density relation for  \mygal, making it impossible to determine $n_e$ with accuracy. Therefore, we have assumed the lower limit of $n_e~\sim$~100~\,cm$^{-3}$ for the galaxy in the following analysis.

\subsubsection{Gas excitation}
\begin{figure*}
	\includegraphics[scale=0.8]{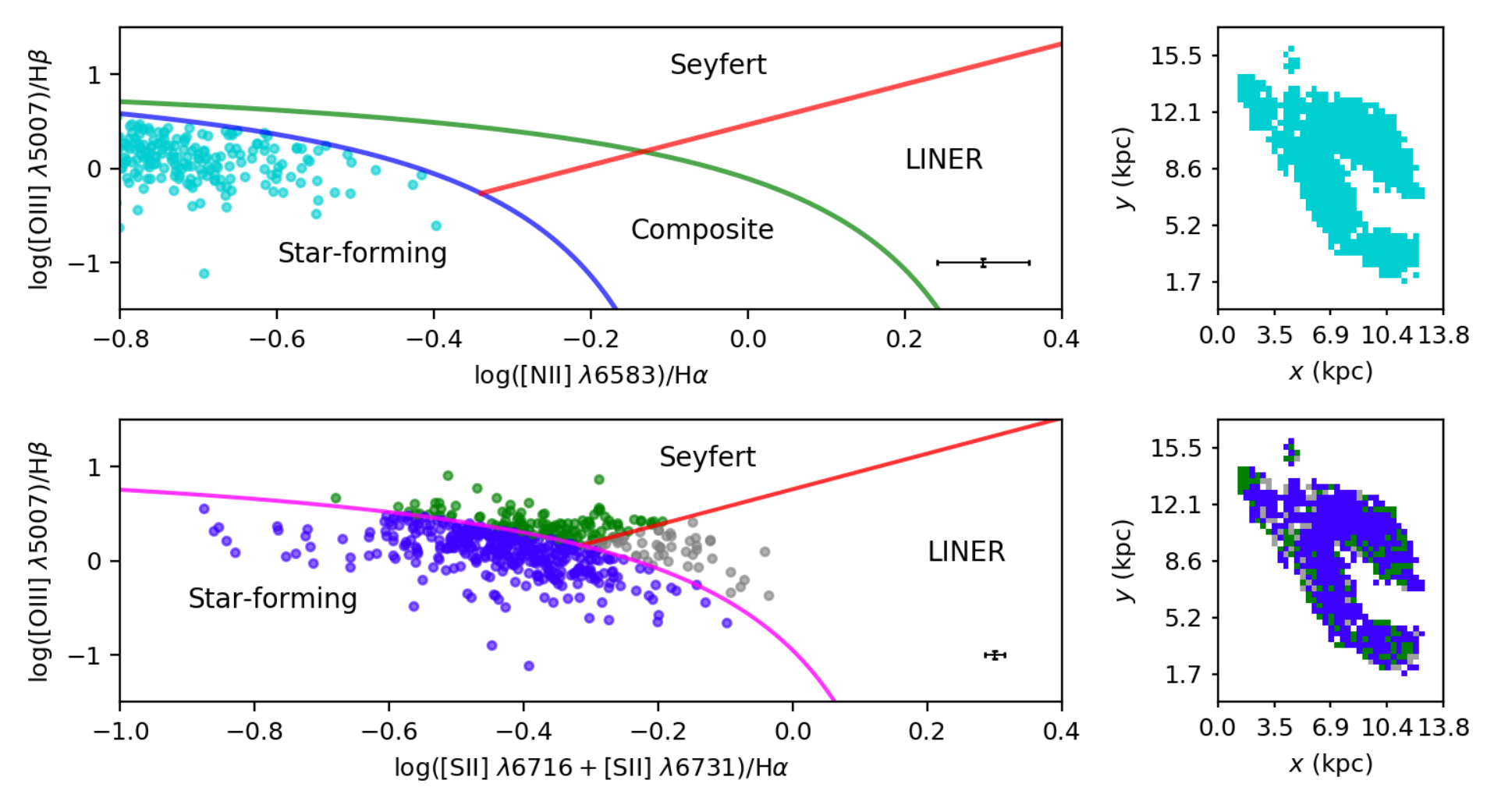}
    \caption{Spatially resolved diagnostic diagram for \mygal. \textit{Left top panel:} The circles represent the position of individual spaxels in the diagnostic diagram. The relations for the solid blue and red lines were taken from \citet{kauffmann2003}. The relation for the solid green line was obtained from \citet{kewley2001}. \textit{Left bottom panel:} Same as left top panel, now using different emission line ratio for the diagram. The red and magenta solid lines are given by \citet{kewley2006,kewley2001}, respectively. \textit{Right top/bottom panel}: The \mygalsimple\ with each spaxel colored corresponding to its position in the diagnostic diagram. The bars in the bottom right corner of the right panels indicate the mean error associated to the position of each point in the diagram.}
    \label{fig:BPT}
\end{figure*}

The emission line ratios allow us to determine the nature of the ionisation source of the gas in \mygal. For this analysis, we used the traditional BPT diagnostic diagram \citep{baldwin1981} to create spatially resolved maps of the likely ionisation sources \citep[for a similar analysis see][]{wylezalek2017,nascimento2019}. In Fig.~\ref{fig:BPT} we show these maps for \mygal. In the diagram involving the lines ratios [\ion{N}{ii}]/H$\alpha$ {\it versus} [\ion{O}{iii}]/H$\beta$ (left top panel) the solid blue line separating \ion{H}{ii} from the transition region, as well as the solid red line separating Seyfert and LINER regions were taken from \citet{kauffmann2003}. The solid green line separating transition region from the AGN region was obtained from \citet{kewley2001}. The symbols are the positions of individual spaxels in the diagnostic diagram and the spatial distribution within the galaxy is shown on the right panel. For the diagram involving [\ion{S}{ii}]/H$\alpha$ {\it versus} [\ion{O}{iii}]/H$\beta$ line ratios (left bottom) the solid magenta line is from \citet{kewley2001}, while the solid red line is from \citet{kewley2006}. 

Both diagrams show that the gas is ionised by young massive stars rather than by an AGN. Ionisation by shocks are investigated using the fast radiative shock models from \citet{allen2008}, adopting solar metallicity, $n_{e} = 100 \ \mathrm{cm}^{-3}$ and varying the values of magnetic field. With an inspection of the curves of emission line ratio vs. shock velocity we observe that the values for the gas in \mygal\ are coherent to shock velocities $ < 100$\,km\,s$^{-1}$. In the regime of such low velocities, it is unlikely for shock ionisation to be the dominant mechanism of photoionisation in \mygal.

\subsection{Extinction}
We have already corrected the observed spectra for the Galactic dust extinction using the \citet{schlegel1998} dust maps, but the intrinsic attenuation of \mygal\ still remains, and correcting for this effect is essential for obtaining many of the important properties concerning the gas components. 

\begin{figure}
	\includegraphics[width=\columnwidth]{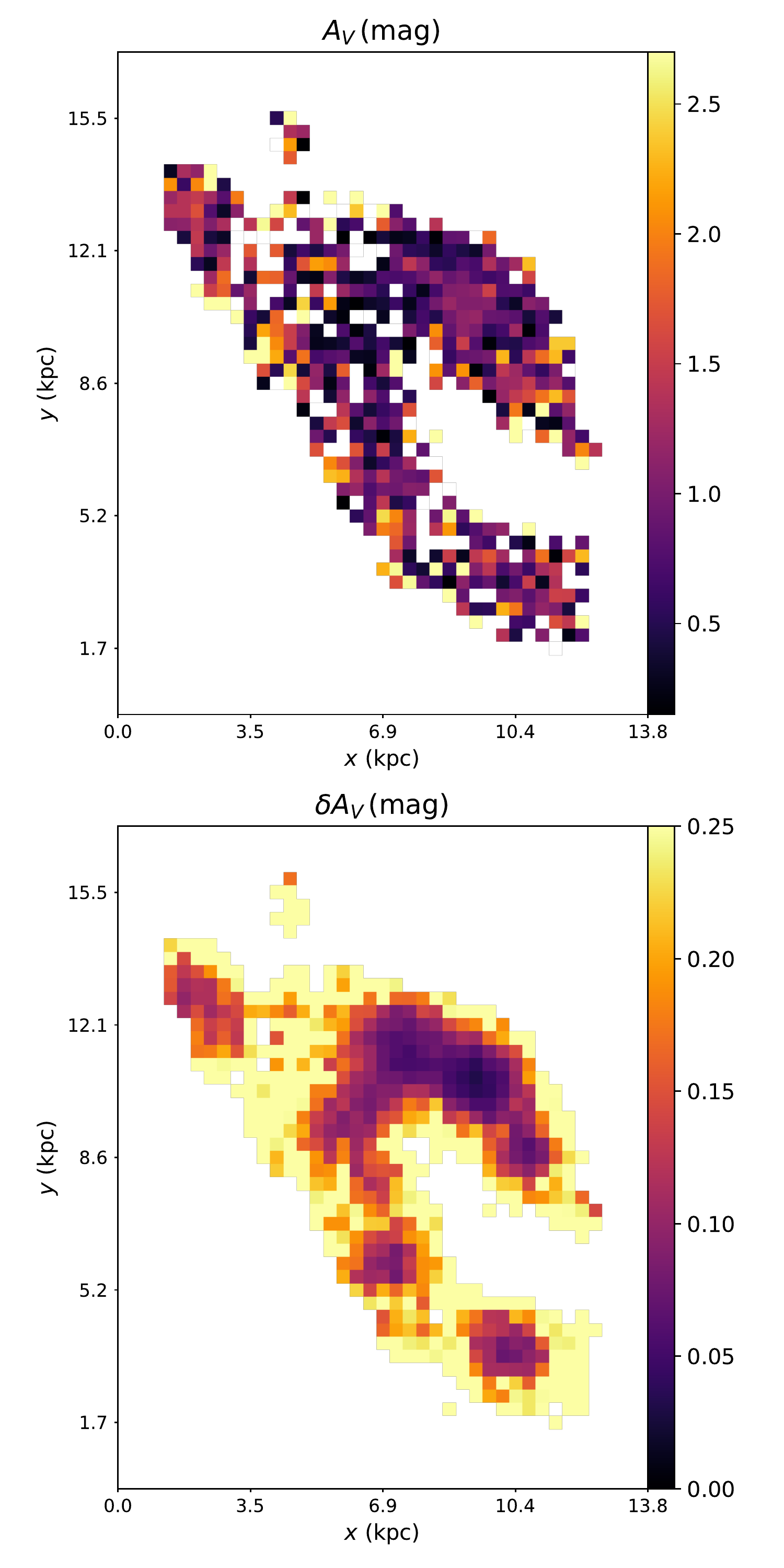}
    \caption{
    \textit{Top Panel:} The extinction map for \mygal. The lighter regions of the map represent higher light extinction, while the darker regions represent lower extinction. In white, the masked points, excluded from the plot. The colour bar shows the range of extinction values measured in the region in magnitude units. \textit{Bottom Panel:} Map with the uncertainties of the $A_{v}$ measurement, represented by $\delta A_{v}$, where we observe that the uncertainties drop as the spaxel is closer to the edge of \mygal.}
    \label{fig:extinction_map}
\end{figure}

To perform this correction we have followed \citet{osterbrock2006}. By considering $I_{\lambda}$ as the observed intensity for a given wavelength and $I_{\lambda_{0}}$ as the intrinsic intensity, we have:

\begin{equation}
    I_{\lambda} = I_{\lambda_{0}} \times 10^{-cf(\lambda)},
    \label{extinction}
\end{equation}
where $c$ is an undetermined constant and $f(\lambda)$ is the extinction curve. The number of magnitudes of extinction, $A_{\lambda}$, is related to the lines intensity ratio by:

\begin{equation}
    A_{\lambda} = -2.5 \ \mathrm{log} \bigg(\frac{I_{\lambda}}{I_{\lambda_{0}}}\bigg).
    \label{alambda}
\end{equation}
$A_{\lambda}$ is related to $A_{V}$ via the reddening law. By assuming the case B of recombination, with an intrinsic intensity ratio of $\mathrm{H} \alpha / \mathrm{H} \beta$ = 2.87, electron temperature of $10^{4}$\,K and and $R_{V} = 3.1$, $A_{V}$ can be derived using

\begin{equation*}
    A_{V} = 7.23 \times \mathrm{log} \bigg[ \frac{F(\mathrm{H} \alpha)}{F(\mathrm{H} \beta)} \bigg] - 3.31,
    \label{eq:reddening}
\end{equation*}
where $F(\mathrm{H} \alpha)$ and $F(\mathrm{H} \beta)$ are the observed fluxes. This equation represents the attenuation caused by dust and molecular gas surrounding the \ion{H}{ii} regions. The extinction map produced with this equation for \mygal\ is shown in Fig.~\ref{fig:extinction_map}.

\subsection{Star formation rate}
\begin{figure}
	\includegraphics[width=\columnwidth]{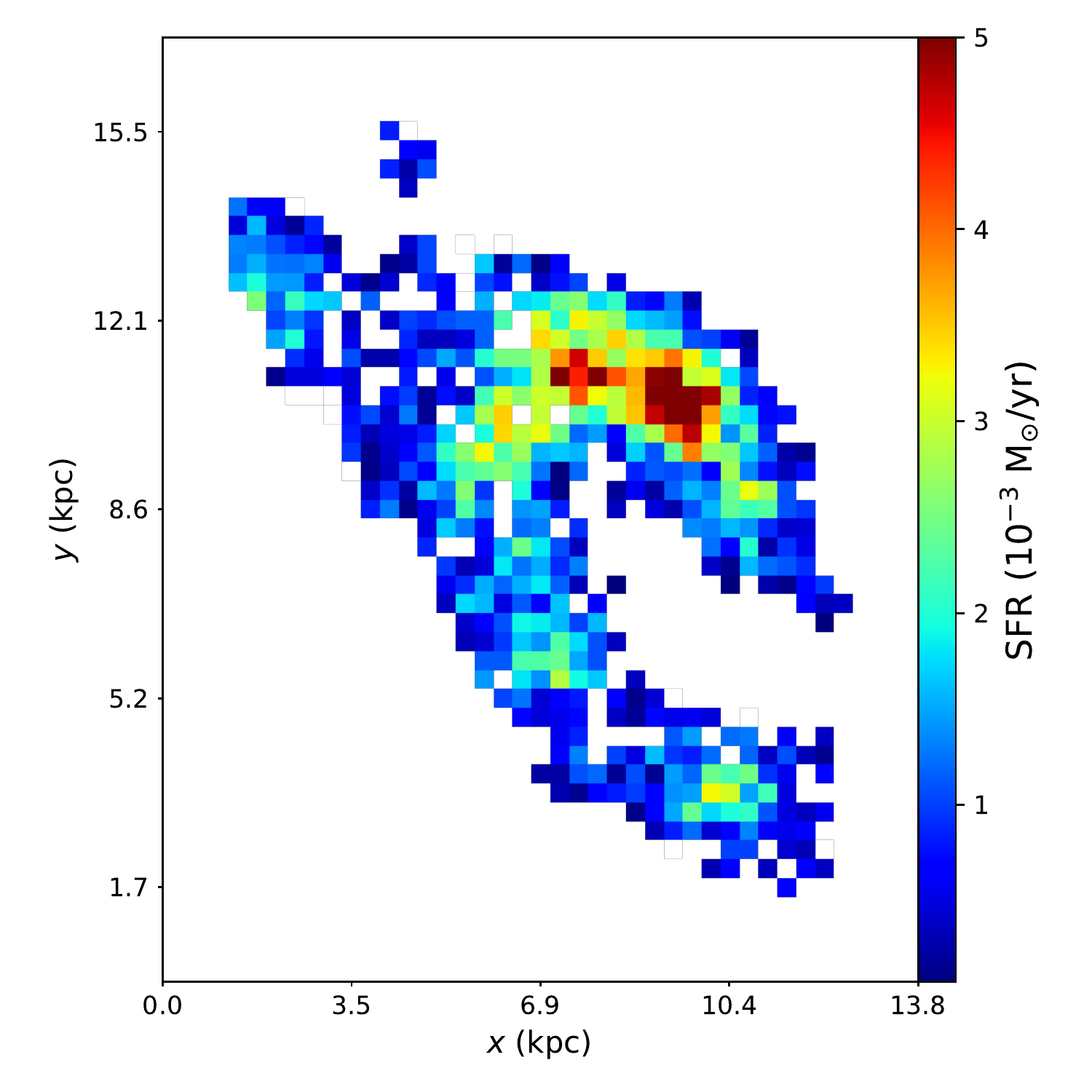}
    \caption{The \mygal\ SFR distribution map. Regions with high SFR appear in red, while regions without significant star formation appear in blue. The colour bar indicates the values of SFR associated to each colour in units of $M_{\odot}$\,yr$^{-1}$.}
    \label{fig:SFR}
\end{figure}

From the diagnostic diagram one can see that all spaxels of \mygal\ are in the star-forming region, so it is interesting to estimate its Star Formation Rate (SFR). We can calculate it using the Balmer recombination lines. Assuming case B of recombination and a Salpeter Initial Mass Function (IMF), we can use the expression below to estimate the instantaneous SFR of \mygal\  \citep{kennicutt1998}:

\begin{equation}
    \mathrm{SFR} (M_{\odot} \ \mathrm{yr}^{-1}) = 7.9 \times 10^{-42} L(\mathrm{H} \alpha) \ (\mathrm{erg} \ \mathrm{s}^{-1}).
    \label{eq:SFR}
\end{equation}
The luminosity $L(\mathrm{H} \alpha)$ was obtained from the flux of H$\alpha$ emission line using the distance of \mygal, which was derived from its redshift of  $z = 0.04025 \pm 0.00003$. Fig.~\ref{fig:SFR} shows the resulting SFR map. The colour bar shows the SFR in units of $10^{-3}$\,$M_{\odot}$\,yr$^{-1}$. Integrating the SFR over all spaxels results in a total SFR of 0.70 
$M_{\odot}$ yr$^{-1}$
and a 
$\Sigma \mathrm{SFR}$ of 
1.4 $ \times 10^{-2} \ M_{\odot}$ yr$^{-1}$ kpc$^{-2}$,
which was obtained by dividing the SFR by the area of each spaxel. The peak value for an individual spaxel reaches $6.75 \times 10^{-3}$\,$M_{\odot}$\,yr$^{-1}$.

\subsection{Stellar mass}
We can use the information acquired with the stellar population synthesis to estimate the current stellar mass 
$M_{\star}$\footnote{\label{foot:manual_starlight}\url{https://minerva.ufsc.br/starlight/files/papers/Manual_StCv04.pdf}}:

\begin{equation}
    M_{\star} (M_{\odot}) = 4\pi d^{2} \times \mathrm{M}^{\mathrm{cor}}_{\mathrm{tot}} \times (3.826 \times 10^{33})^{-1} \quad,
    \label{eq:stellar_mass}
\end{equation}
where $d$ is the distance to the galaxy in units of cm and the base spectra used is in a proper unit of $L_{\odot}\,$\AA$^{-1}\,M_{\odot}^{-1}$ and the observed spectra in units of erg\,s$^{-1}\,$cm$^{-2}$\,\AA$^{-1}$. $\mathrm{M}^{\mathrm{cor}}_{\mathrm{tot}}$ is a {\sc starlight} output parameter in units of M$_{\odot}$\,erg$^{-1}$\,cm$^{-2}$, which gives the current mass of stars in the galaxy based on the contribution of each SSP to the best-fit synthetic spectrum. We can apply equation \ref{eq:stellar_mass} to each spaxel of our datacube to obtain the total stellar mass for both Mrk~1172 and \mygal. We obtain $M_{\star} \sim 1.2 \times 10^{11} M_{\odot}$ for Mrk 1172 and $M_{\star} \sim 3.0 \times 10^{9} M_{\odot}$ for \mygal. Similarly, we can estimate how many solar masses have been processed into stars through the lifetime of the system ($M^{\mathrm{ini}}_{\star}$)\textsuperscript{\ref{foot:manual_starlight}} by using:

\begin{equation}
    M^{\mathrm{ini}}_{\star} \ (M_{\odot}) = 4\pi d^{2} \times \mathrm{M}^{ini}_{\mathrm{tot}} \times (3.826 \times 10^{33})^{-1} \quad ,
    \label{eq:processed_mass}
\end{equation}
where $\mathrm{M}^{ini}_{\mathrm{tot}}$ represents the mass that has been converted into stars and is given in units of M$_{\odot}$\,erg$^{-1}$\,cm$^{-2}$\citep{riffel2021} \textsuperscript{\ref{foot:manual_starlight}}. Applying equation \ref{eq:processed_mass} to all valid spaxels, we obtain $M^{\mathrm{ini}}_{\star} \sim 1.7 \times 10^{11} M_{\odot}$ for Mrk 1172 and $M^{\mathrm{ini}}_{\star} \sim 3.9 \times 10^{9} M_{\odot}$ for \mygal. $M^{\mathrm{ini}}_{\star}$ is expected to be higher than the current stellar mass of the system due to the mass returned to the ISM by SNe and winds.

\subsection{Ionised gas mass}
Following, for example, \citet{nascimento2019}, we can calculate the mass of the ionised gas by using the expression:

\begin{equation}
    \mathrm{M} = n_{e} m_{p} V f,
    \label{ionized_gas_mass}
\end{equation}
where $n_{e}$ is the electron density of the gas, $m_{p}$ is the proton mass, $\mathrm{V}$ is the volume of the ionised region and $f$ is the filling factor. Using the emissivity of H$\beta$ ($j_{\mathrm{H}\beta}$), we can calculate the total luminosity of this line:

\begin{equation}
    \mathrm{L}(\mathrm{H} \beta) = \int \int j_{\mathrm{H}\beta} \ \mathrm{d}\Omega \mathrm{dV}.
    \label{integral_LHb}
\end{equation}
We know from \citet{osterbrock2006} that for recombination case B (in the low-density limit), assuming $\mathrm{T} = 10^{4}$\,K we have:

\begin{equation*}
    \frac{4\pi j_{\mathrm{H} \beta}}{n_{e} n_{p}}=1.24 \times 10^{-25} \ \frac{\mathrm{erg} \ \mathrm{cm}^{3}}{\mathrm{s}},
\end{equation*}
where $n_{e}$ and $n_{p}$ are the the electric and proton densities, respectively. Using this result into the integral in equation \ref{integral_LHb} we obtain $\mathrm{L} (\mathrm{H} \beta)$ in units of erg s$^{-1}$:

\begin{equation*}
    \mathrm{L}(\mathrm{H}\beta) = 1.23 \times 10^{-25} \ n_{e} n_{p} Vf.
\end{equation*}
Assuming the gas is completely ionised ($n_{e} = n_{p} = n$) we can isolate $Vf$ from the expression above and use it in equation \ref{ionized_gas_mass}. Considering the lower limit of $n_{e} = 100 \ \mathrm{cm}^{-3}$, we obtained the mass of the ionised gas:

\begin{equation}
    M(M_{\odot}) = 6.782 \times 10^{-35} L(\mathrm{H} \beta),
    \label{eq:ionized_gas_mass}
\end{equation}
where $L(\mathrm{H} \beta)$ is the luminosity of H$\beta$ emission line, in units of ergs\,s$^{-1}$. Since we used $n_{e} = 100 \ \mathrm{cm}^{-3}$, the mass resulting from equation \ref{eq:ionized_gas_mass} should be interpreted as a lower-limit mass. We applied equation \ref{eq:ionized_gas_mass} to \mygal, where H$\beta$ is strong for the majority of the spaxels in the region, by using the reddening corrected $F(\mathrm{H} \beta)$ to calculate $L(\mathrm{H} \beta)$ for each spaxel in \mygal. Integrating over the whole region corresponding to \mygal, we have obtained $M = 3.8 \times 10^{5} M_{\odot}$ for the ionised gas mass. 
\subsection{Oxygen Abundances}
We aim to characterise \mygal\ with respect to metallicity, using as proxy the Oxygen abundance in its ISM.
Since we are not able to directly measure the electron temperature ($T_e$) because the [\ion{O}{iii}]\,$\lambda$4363\,\AA\ emission line is out of our observed region we cannot use the direct method \citep[e.g.][]{osterbrock2006}. However, many calibrations were made in the past decades without the need of $T_e$ determination  \citep[e.g.][known as indirect methods]{pettini2004,nagao2006,perez2009,marino2013}. Useful calibrations are given in \citet{marino2013}:

\begin{align}
\mathrm{log (O/H)} + 12 &= 8.533 (\pm 0.012) -0.214(\pm 0.012) \times O3N2\\ 
\mathrm{log (O/H)} + 12 &= 8.743 (\pm 0.027) + 0.462(\pm 0.024) \times N2,
\end{align}

\noindent where $O3N2$ and $N2$ indices are defined as follows \citep{alloin1979,thaisa1994}:

\begin{align}
 O3N2 &= \mathrm{log} \bigg(\frac{[\mathrm{O}\,\mathrm{III}]\lambda 5007}{\mathrm{H}\beta} \times \frac{\mathrm{H}\alpha}{[\mathrm{N}\,\mathrm{II}]\lambda 6583} \bigg) \\ 
 N2 &= \mathrm{log} \bigg( \frac{[\mathrm{N}\,\mathrm{II}]\lambda 6583}{\mathrm{H}\alpha}  \bigg),
 \label{eq:abundance}
\end{align}

The ionisation parameter in galaxies between $0 < z < 0.6$ changes such that the line ratios [\ion{O}{iii}]/H$\beta$ and [\ion{N}{ii}]/H$\alpha$ from giant low surface brightness \ion{H}{ii} regions begin to rise and lower, respectively. When using a metallicity tracer for galaxies in that redshift range, is crucial either to take into consideration the behaviour of the ionisation parameter and the change in the emission line ratios or to use a tracer that is independent of the conditions of ionisation of the ISM \citep{Monreal_Ibero2011,Kewley2015}. With this in mind, we used a calibration based on [\ion{N}{ii}] and [\ion{S}{ii}] in addition to the previous calibrations, given by \citet{dopita2016}:

\begin{equation}
    \mathrm{log (O/H)} + 12 = 8.77 + \mathrm{log ([\ion{N}{ii}]/[\ion{S}{ii}])} + 0.264 \times \mathrm{log([\ion{N}{ii}]/H\alpha)}
    \label{eq:abundance_dopita}
\end{equation}

\begin{figure*}
	\includegraphics[scale = 0.27]{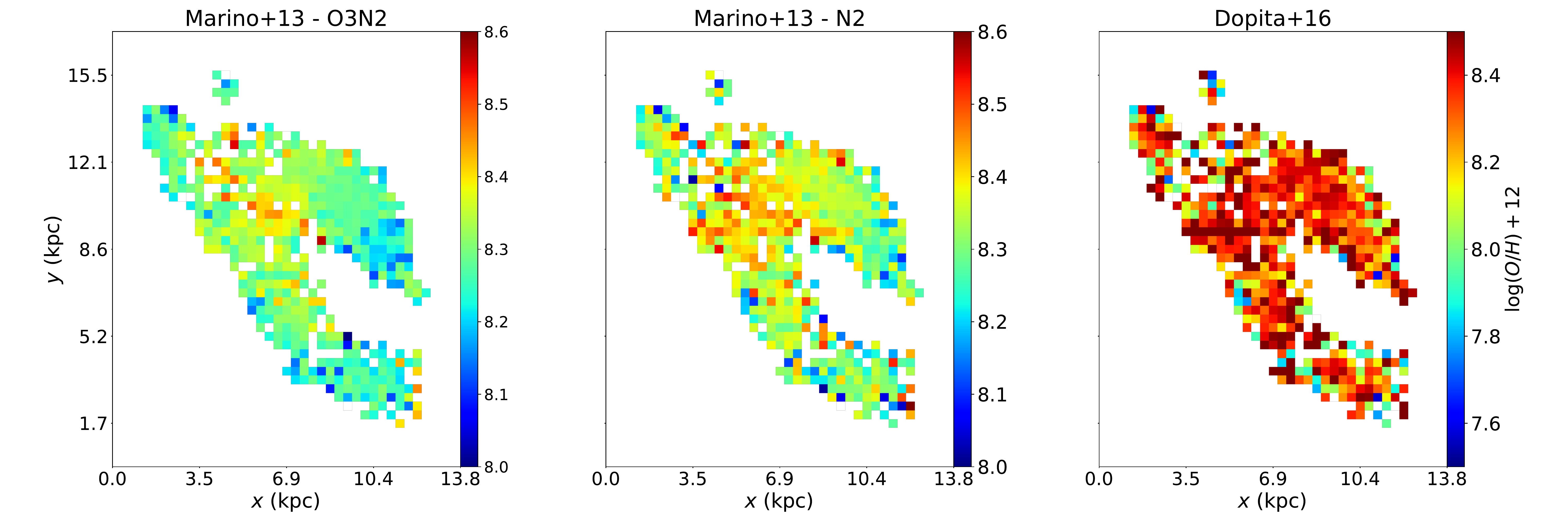}
    \caption{
    Maps for Oxygen abundance obtained using the indirect method, in units of $\mathrm{log} (O/H) + 12$. In left and central panels, the results using O3N2 and M2 indexes from \citet{marino2013}. In right panel, the result using the calibration from \citet{dopita2016}.
    }
    \label{fig:abundances}
\end{figure*}

These calibrations were applied to each spaxel of \mygal, producing three Oxygen abundance maps, two for each tracer of \citet{marino2013} and the third for the calibration from \citet{dopita2016}. These are shown in Fig.~\ref{fig:abundances}. In the first two cases we obtain values in the range of $8.0 < \mathrm{log} (O/H) + 12 < 8.6$, which represents a range of approximately $0.2\,Z_{\odot} < Z < 0.7\,Z_{\odot}$. In the third case, the metallicity spans a wider range of $7.5 < \mathrm{log} (O/H) + 12 < 8.5$ (i.e., $0.05\,Z_{\odot} < Z < 0.5\,Z_{\odot}$). To have a representative value, we calculated the average metallicity for each map. We have obtained $\mathrm{log} (O/H) + 12 = 8.28 \pm 0.03$ for the O3N2 index and $\mathrm{log} (O/H) + 12 = 8.296 \pm 0.004$ for the N2 index. For the calibration given in \citet{dopita2016} we obtained $\mathrm{log} (O/H) + 12 = 8.2 \pm 0.2$. As a sanity check we have calculated the abundance derived from the emission lines measured in an integrated spectrum of \mygal, and the values are consistent.

\subsection{Kinematics}
\begin{figure*}
	\includegraphics[scale=0.73]{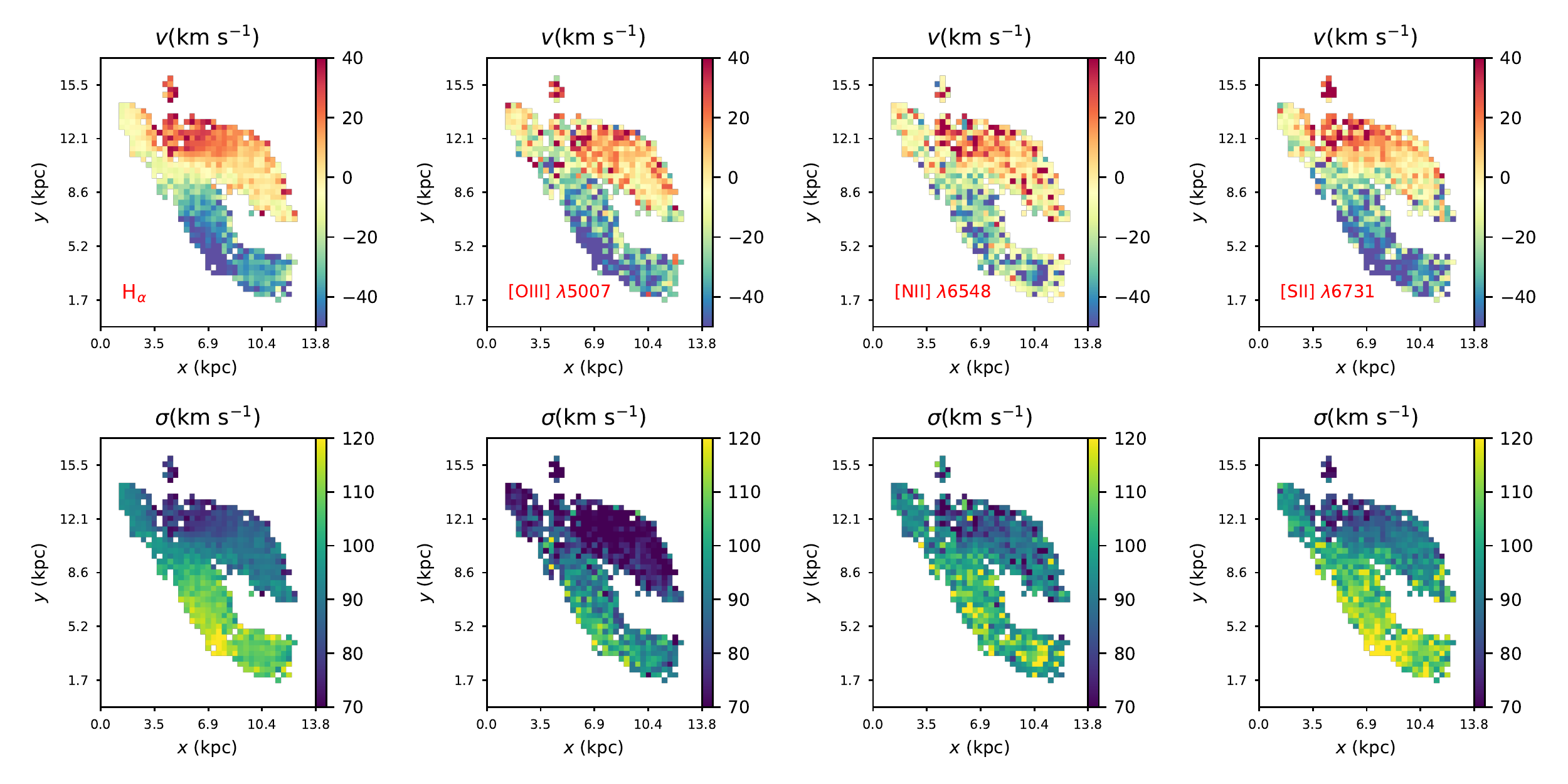}
    \caption{Kinematic maps for H$ \alpha$, [\ion{O}{iii}]$\lambda \ 5007$, [\ion{N}{ii}]$\lambda \ 6548$ and [\ion{S}{ii}]$\lambda \ 6731$ emission lines. In the upper panel, radial velocity maps, and in the lower panel, the turbulence of the gas. Both parameters are obtained from a single gaussian used to fit the observed emission lines using {\sc ifscube}.
    }
    \label{fig:kinematics}
\end{figure*}

There are many intriguing aspects of \mygal: its irregular morphology, its proximity to Mrk~1172 and its physical properties that indicate it may be interacting with Mrk~1172. Thus it is important to explore the kinematics of \mygal. We can obtain the radial velocity, $v$, of the gas for the identified emission lines in \mygal\ spectra directly from {\sc ifscube} single gaussian fit, as well as the velocity dispersion, $\sigma$. In the fit process, we fixed the doublets as being in the same kinematic group, meaning that the resulting velocity values will be the same for the same ion. In Fig.~\ref{fig:kinematics}, we show both velocity, corrected by subtracting off the mean velocity of the system, and velocity dispersion ($\sigma$) maps for H$\alpha$, [\ion{O}{iii}]$\lambda 5007$, [\ion{N}{ii}]$\lambda 6548$ and [\ion{S}{ii}]$\lambda 6731$ emission lines.

First we observe, especially in the maps of centroid velocity, that the gradient seen is more smooth in the case of H${\alpha}$ emission line maps. This is due to the fact that in spaxels near the edge of \mygal\ (which was defined  using the strongest emission line in its spectra, i.e., H$\alpha$) the SNR is low and the measurement of emission lines other than H$\alpha$ have higher uncertainties. Even so, all velocity maps exhibit the same trend for the motion of the gas, where the upper part of the galaxy is moving away from us while the bottom is approaching. Since we do not know the distances with precision we can not determine the orientation of the motion of the dwarf galaxy around the ETG.

The maps of velocity dispersion of the gas seem to replicate the trend seen in the maps of radial velocity. In the case of the kinematical group corresponding to the emission lines of [\ion{O}{iii}] we observe that more spaxels reach values of $\sigma < 90$\,km\,s$^{-1}$ in comparison to the other maps. Such a difference could be caused, for instance, if this zone of ionisation is closer to the region with higher flux of ionising photons in comparison to the other zones of ionisation and being strongly affected by the winds of massive stars. However this difference rarely becomes larger than 20\,km\,s$^{-1}$, as can be observed in Fig.~\ref{fig:kinematics}, and the most likely is that it is not significant. To determine the zero point in the velocity scale of Fig.~\ref{fig:kinematics}, we used the rest-frame velocity, calculated using the integrated spectrum of \mygal. The values of velocity dispersion were corrected by instrumental width, assuming a resolving power $R$ of 1750 at $\lambda \sim$\,5000\,\AA\ and $R \approx 2500$ at $\lambda \sim$\,6500\,\AA.

\begin{figure}
    \centering
    \includegraphics[scale = 0.58]{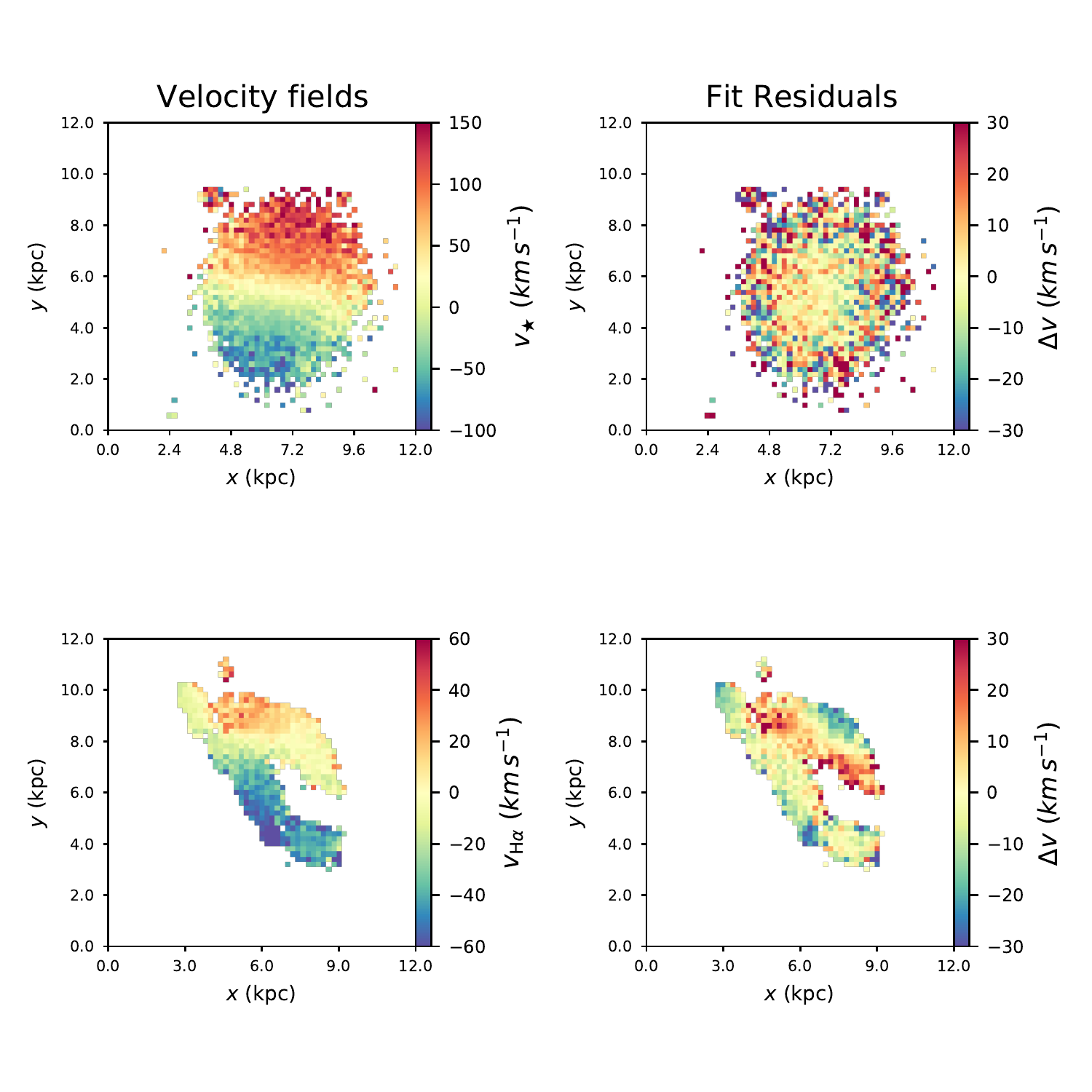}
    \caption{\textit{Top left panel:} Stellar velocity field of Mrk~1172. \textit{Bottom left panel:} Velocity field for the gas in \mygal\, measured using H$\alpha$ emission line. The values of velocity were obtained by subtracting the systemic velocity of the galaxy ($v_{s}$ = -240 km\,s$^{-1}$) from the centroid velocity obtained previously. \textit{Right panels:} Residual maps for the model fitted.}
    \label{fig:kinematical_model}
\end{figure}

Although we do not have information about the gas of Mrk 1172 we can explore its stellar kinematics using the results obtained in the stellar population synthesis. In Fig.~\ref{fig:kinematical_model} we present, in the top left panel, the stellar velocity field for the ETG. To better understand the interaction between both galaxies we used an analytical model which assumes that the gas has circular orbits around the plane of a disk \citep{kruit1978,bertola1991}. For this model we used the stellar velocity field of Mrk~1172 and the gas velocity field measured from H$\alpha$ shown in Fig. \ref{fig:kinematics} and shown in bottom left panel of Fig.~\ref{fig:kinematical_model}. The equation that gives the model velocity field is given by:

\begin{align*}
&V(R,\Psi) =  v_{s}  +  \\ 
& \frac{A R \cos(\Psi - \Psi_{0}) \sin(i)\cos^{p}(i)}{\{R^{2} [\ \sin^{2}(\Psi - \Psi_{0}) + \cos^{2}(i)\ \cos^{2}(\Psi - \Psi_{0})] + C_{0}^{2} \cos^{2}(i)  \}^{p/2}} \ ,
\end{align*}
where $R$ is the radial distance to the nucleus projected in the plane of the sky with a corresponding position angle $\Psi$, $v_{s}$ is the systemic velocity of the analysed galaxy, $A$ is the velocity amplitude, $\Psi_{0}$ is the position angle of the line of nodes, $C_{0}$ is a concentration parameter defined as the radius where the rotation curve reaches 70\% of the velocity amplitude and $i$ is the disc inclination in relation to the plane of the sky. The parameter $p$ varies from 1.0 to 1.5.
In table \ref{tab:fit_parameters} we summarize the results obtained for the fit. The values of $v_s$ presented were obtained by adding the heliocentric velocity of each galaxy. The residual maps of the fit are presented in the right panel of Fig.~\ref{fig:kinematical_model}

\begin{table}
\centering
\caption{Final result of the fitting process for both galaxies.}
\begin{tabular}[!t]{lcc}
\hline
                                      & Mrk 1172                   & \mygal      \\ \hline
$v_{s}$ (km\,s$^{-1}$) & $12373 \pm 5$    & $11835 \pm 3$  \\
$A$ (km\,s$^{-1}$)     & $720 \pm 180$ & $310 \pm 97$  \\
$i$ ($^{\circ}$)                      & $30 \pm 9$    & $28 \pm 10 $  \\
$C_{0}$ (arcsec)                      & $ 4.8 \pm 0.5$ & $4.0 \pm 0.9$ \\
$\Psi_{0}$ ($^{\circ}$)               & $167 \pm 4$   & $165 \pm 4$   \\
$p$                                   & 1.5                        & 1.5                        \\ \hline
\end{tabular}
\label{tab:fit_parameters}
\end{table}

\section{Discussion}\label{sec:4}
One important question that arises at this point is: \textit{what is the origin and nature of the  nebular emission region identified throughout this paper as \mygal\ and how is it connected to ETG Mrk~1172?} In this section we will use the results presented previously to build possible scenarios for the observed system.

From the stellar population analysis it is clear that Mrk~1172 and \mygal\ do not share similar SFHs. While the ETG is dominated by an old stellar population with no signs of ionised gas, \mygal\ presents a rich emission line spectra with its stellar emission dominated by very young stellar populations ($t < 10^{9}$, see Figs.~\ref{fig:synth_results} and \ref{fig:mean_age}). It is worthwhile mentioning that although Fig.~\ref{fig:synth_results} displays information for individual spaxels, the results are representative of all the spaxels in each system, and arguments based on this figure can be extended to the entirety of both systems. In Fig.~\ref{fig:synth_results} we also observe a significant contribution of very young stellar populations ($t < 10^{8}$\,yrs) in the spectra of Mrk~1172. These are probably artificial stellar populations caused by i) the presence of an AGN in Mrk~1172 or (ii) the absence of blue horizontal branch stars in the models, forcing a young population to take into account the blue light of these stars in the ETG \citep{cid2010,delgado2010}. 

An interesting parameter to explore is the magnitude in the B-band, since it probes the young stellar populations in the galaxy. Although we lack the spectral range to measure the B-band magnitude by integrating the spectra, we can use the photometric values of SDSS to convert to B band using \citep{jordi2006}:     
\begin{equation}
    B - g = (0.349 \pm 0.009) \ (g-r) + (0.245 \pm 0.006).
    \label{eq:mag_conv}
\end{equation}

From SDSS we have the magnitudes of \mygalsimple:  $u = 22.81 \pm 0.52$, $g = 20.65 \pm 0.04$, $r = 20.19 \pm 0.04$, $i = 19.70 \pm 0.04$ and $z = 19.69 \pm 0.11$. By applying equation \ref{eq:mag_conv} we obtain $B = 21.00 \pm 0.37$, which corresponds to an absolute magnitude of $M_{B} \sim -15$ mag, which means the observed dwarf galaxy is faint in the blue band.

We also wish to understand the mechanism responsible for the gas excitation, evidenced by the strong emission lines observed in \mygal\ spectra. 
From the BPT diagnostic diagrams shown in Fig.~\ref{fig:BPT} it is clear that the gas within \mygal\ is excited by young hot stars rather than by a hard radiation field. The few spaxels falling in the Seyfert region of the diagnostic diagram are located at the edges of the galaxy and have low SNR, as indicated by the transparency of the circles. We compared our line ratios with predictions of shock models from \citet{allen2008} and concluded that fast radiative shocks are unlikely to be a significant mechanism of ionisation of the gas in \mygal.

Due to the presence of young stellar populations and the excitation mechanism of the gas, we can say that \mygal\ is forming stars, although the efficiency of the process is unknown. The young stars that are ionising the gas seem to be located in a few clumpy knots along the structure of \mygal, better traced by the SFR map in Fig.~\ref{fig:SFR}. This structure also becomes visible when looking at the H$\alpha$ collapsed image in Fig.~\ref{fig:Figura1} and in the continuum H$\alpha$ emission in Fig.~\ref{fig:2d_mask}, and are typical structures formed in regions with active star formation. Using the Balmer decrement we produced a reddening map for \mygal, shown in Fig.~\ref{fig:extinction_map}. The mean $A_{V}$ is 0.9\,mag, but in few spaxels this value is up to $\sim$\,2.0\,mag, reinforcing the inhomogeneity of the medium, or also the uncertainty in the emission line ratio for these spaxels. The intensities of the emission lines, used through this paper, have been corrected for dust attenuation. 

Using the emission line fluxes we calculated the instantaneous SFR using equation \ref{eq:SFR}, obtaining a SFR of 0.70 M$_{\odot}$ yr$^{-1}$. We also derived the spatial distribution of SFR for \mygal\ ($\sum_{\mathrm{SFR}}$), resulting in $1.4 \times 10^{-2} \ \mathrm{M}_{\odot} \mathrm{yr}^{-1} \mathrm{kpc}^{-2}$. Finally, we calculated a lower limit for the mass of ionised gas in \mygal\ using the H$\beta$ emission line strength, resulting in a mass of $3.8 \times 10^{5} \mathrm{M}_{\odot}$.  

From the stellar population synthesis we  obtained an estimate for the stellar mass of both galaxies. We estimate the values of $3.0 \times 10^{9}$\,$M_{\odot}$ for \mygal\ and $1.2 \times 10^{11}$\,$M_{\odot}$ for Mrk~1172. In the case of the latter, the order of magnitude of this value is coherent to what is expected for massive ETG like Mrk~1172, and is in agreement with a previous estimates for the stellar mass for this galaxy \citep[$8.91 \times 10^{10}$\,$M_{\odot}$,][]{omand2014}.  On the other hand, \mygal\ has an estimated value of stellar mass comparable to dwarf galaxies, like the SMC, for example \citep[$6.5 \times 10^{9}$\,$\mathrm{M}_{\odot}$,][]{bekki2009}.

With the values of stellar mass and SFR, we can estimate the specific Star Formation Rate (sSFR) for \mygal, which can be compared with a previous value found in the catalogue of \citet[][DV19 throughout this paper]{delli2019}, where the sSFR is measured using photometric information for a large sample of galaxies from SDSS DR7. We have obtained a value of log(sSFR)$= -9.63$, which is considerably different than the log(sSFR)$= -10.85$ found in DV19. However, these two values were eerived using different techniques, and the photometry of \mygal\ is flagged as medium quality in DV19, meaning the direct comparison between both values is hard to perform.

As observed before in this work, \mygal\ has an irregular shape and lies next in projection to Mrk~1172. Thus, a natural first assumption is that both galaxies are interacting, and the irregular shape of \mygal\ could be partially explained by tidal forces. Analysing the available redshift for Mrk~1172 (which is consistent with our determination) and our measurement for \mygal, one can calculate a distance of approximately 4\,Mpc between both galaxies, for which gravitational interactions could be neglected. However, this calculation does not take into account that the shift in the emission lines of \mygal\ could be also caused by its motion in relation to Mrk~1172. Considering that Mrk~1172 has a massive halo, it is possible that this halo is populated by faint dwarf galaxies. 
In this case, the participation of \mygal\ in this group of galaxies should not be discarded, and the estimate of 4\,Mpc of distance between both galaxies could be incorrect. As an exercise, if we assume 300\,km\,s$^{-1}$ for the group velocity, for example, it would affect the above mentioned distance estimate by approximately 3\,Mpc, i.e., the uncertainty in the distance estimate due to the relative motion of the dwarf galaxy is large. 

To investigate deeper the interaction between both galaxies, we fit a model to the stellar velocity field of Mrk~1172 and to the H$\alpha$ velocity field of \mygal~(see Fig.~\ref{fig:kinematical_model}). The final results of the fit are listed in table \ref{tab:fit_parameters}. It can be seen that both galaxies present similar results for the disc inclination and the position angle of the line of the nodes ($i$ and $\Psi_{0}$, respectively). Although the presence of a disc in \mygal\ is uncertain, the model fit is satisfactory, and this result indicates that the motion of the gas in the dwarf galaxy is affected by the ETG. These considerations, combined to the irregular shape of the dwarf galaxy, leads us to the conclusion that Mrk~1172 and \mygal\ are currently interacting.

Finding a low metallicity system next to an ETG, as seems to be the case for Mrk~1172 and \mygal, deserves further investigation.
Here we use Oxygen abundance as a tracer for the metallicity of the ISM of \mygal\ and we employ the calibrations from \citet{marino2013} for each spaxel of \mygal. The average metallicity obtained is $\mathrm{log} (O/H) + 12 = 8.28 \pm 0.03$ for the O3N2 index and $\mathrm{log} (O/H) + 12 = 8.296 \pm 0.004$ for the N2 index. Since O3N2 is based in emission lines both in the blue and red parts of the spectrum and suffers a strong dependency on the ionisation parameters for the ISM ionisation conditions of \mygal. Thus we used the calibration from \citet{dopita2016}, which has no dependency on the ionisation parameters of the galaxy, to compare with the previous results. Using this calibration we obtained an average value of $\mathrm{log} (O/H) + 12 = 8.2 \pm 0.2$. In general, this third calibration gives slightly lower values of metallicity in comparison to the other maps. Combining the values obtained it is possible to estimate that the metallicity of \mygal is approximately 1/3\,$Z_{\odot}$. The range of metallicites observed in \mygal is also observed in metal-poor systems such as dwarf irregulars and some BCDGs \citep{kunth2000,gildepaz2003}. This feature raises another question: \textit{what is the origin of this metal-deficient content?}

Models of galaxy formation in the $\Lambda$-CDM context predict the inflow of metal-deficient gas from the cosmic web. Numerical simulations indicate that this gas can trigger star formation in disc galaxies and dwarf galaxies of the Local Universe \citep{sanchezalmeida2015}. The triggering of a starburst may happen as the infalling gas gets compressed when it approaches the disc or also it may be accreted and build up the gas mass of the galaxy to eventually give rise to the starburst due to internal instabilities \citep{dekel2009,ceverino2016}. In many cases chemical inhomogeneities can be observed in galaxies in the Local Universe, where localized kpc-size starbursts present considerably lower abundances in comparison to the surrounding ISM. The scenario described above is a plausible interpretation for the metallicity drop in these starburst regions. We believe that the origin of the metal deficient gas in \mygal\ could be caused by infalling gas from the cosmic web, but if this is true, should we not observe a gradient in metallicity in Fig.~\ref{fig:abundances} instead of the homogeneous distribution that is actually observed? As we have seen, the gas in \mygal\ is subject to the interaction with Mrk~1172 and to the inner kinematics of the galaxy. In a scenario of accretion is possible that this metal deficient content has mixed with the ISM.

Therefore our \textit{conclusions about the nature of this object is that it is a dwarf irregular galaxy interacting with the massive ETG Mrk~1172}. It contains young stellar populations that have formed recently, and shows evidence of ongoing star formation. The star formation seems to be taking place mainly in clumpy knots along the structure of the galaxy. The gas phase of the galaxy is metal-poor, which could be related to infalling gas from the cosmic web, although we lack the information to support this hypothesis. To further investigate this hypothesis, observations of cold molecular gas would be needed, and, to the best of our knowledge, so far these observations are not available.  
\section{Summary and conclusions}\label{sec:5}
In this work we present a characterisation of chemical and physical properties of a dwarf galaxy (SDSS~J020536.84-081424.7; \mygalsimple) located approximately ${2.5}''$ from the ETG Mrk~1172. To the best of our knowledge the spectrum of this galaxy was not presented in the literature before, thus the analysis of this work is unprecedented. Below, we summarize the main conclusions of this work:

\begin{itemize}
 
 \item Despite its low SNR in the continuum, the spectra of \mygal\ presents strong emission lines, namely: $\mathrm{H} \alpha$, $\mathrm{H} \beta$, [O {\sc iii}]~$\lambda \lambda$4959+5007, [N {\sc ii}]~$\lambda \lambda$6548+6583 and [S {\sc ii}]~$\lambda \lambda$6716+6731. The stellar population synthesis reveals that Mrk~1172 is predominantly dominated by old stellar populations with $t \geq 10^{10}$\,yrs, while \mygal\ contains young to intermediate age stellar populations, with ages of $10^{7} \sim 10^{8}$\,yrs and $10^{9}$\,yrs. From analysis of  the light weighed stellar populations we conclude that both galaxies have very different SFHs.
 
 \item From the stellar content of \mygal\ and Mrk~1172 we are able to estimate the stellar mass of both galaxies. We obtain $M_{\star} \sim 1.2 \times 10^{11} M_{\odot}$ for Mrk~1172, a value that lies close to a previous measurement found in the literature \citep{omand2014}. The  \mygalsimple\ presents a stellar mass of $M_{\star} \sim 3 \times 10^{9} M_{\odot}$, which is comparable to the mass of dwarf galaxies like the SMC, \citep[$M_{\star}$ = 6.5 $\times\,10^{9}\,\mathrm{M}_{\odot}$][]{bekki2009} and to the BCDG Henize 2-10 \citep[$M_{\star} = 3.7 \times 10^{9} \mathrm{M}_{\odot}$][]{reines2011}, for example. We have also estimated the lower limit of $3.8 \times 10^{5} M_{\odot}$ for the ionised gas mass.
 
\item In order to investigate the gas excitation mechanism, we perform a spatially resolved study making use of emission line diagnostic diagrams such as the BPT. Both diagrams indicate that the gas within \mygal\ is being photoionized by young massive hot stars rather than by an AGN. Shock models are also investigated, but they are negligible.
 \item The metallicity of the ionised gas phase of \mygal\ is $\mathrm{log} (O/H) + 12 = 8.28 \pm 0.03$ for the O3N2 index, $\mathrm{log} (O/H) + 12 = 8.296 \pm 0.004$ for the N2 index and $\mathrm{log} (O/H) + 12 = 8.2 \pm 0.2$ when considering the calibration independent of ionisation parameters from \citet{dopita2016}. These values represent an overall metallicity of approximately 1/3\,$Z_{\odot}$ for \mygal. This galaxy can be considered metal deficient, presenting typical values for dIs.
 \item Our measurements and observations strongly suggest that \mygal\ is actively forming stars. We observe clumpy knots in  $\mathrm{H}\alpha$ emission along the structure of \mygal, which we interpret as being sites of active star formation. We obtain an integrated value of SFR = 0.70 \,$M_{\odot}$\,yr$^{-1}$;  $\sum_{\mathrm{SFR}}$ = 1.4 $ \times 10^{-2} \ M_{\odot}$\,yr$^{-1}$\,kpc$^{-2}$, and a log(sSFR) = -9.63. The latter differs significantly from a previous measurement of log(sSFR) = -10.85 found in the literature (DV19). We attribute this difference to the quality of photometric data for \mygal (flagged as medium in DV19) and the distinct procedures adopted in both works.
 \item Radial velocity maps indicate a trend in the motion of the gas, which seems to be in rotation. Using the velocity field considering H$\alpha$ and the stellar velocity field of Mrk~1172 we fit a kinematic model that considers a gas orbiting in a plane of a disk. From this fit we observe that the position angle of the line of the nodes and the disc inclination are similar for both galaxies, thus indicating that the motion of the gas is being affected by the ETG. We thus conclude that both galaxies are interacting.
\end{itemize}

From the  results summarized above we conclude that the faint nebular emission line region close in projection to Mrk~1172 is actually a gas rich low-metallicity dwarf galaxy, most likely a dwarf irregular. The origin of this galaxy is still uncertain, and we are still unsure as to which specific processes triggered the recent star formation episodes in \mygal.
Bright knots are readily visible in the H$\alpha$ images of \mygal, where the highest fraction of its stellar content seems to be located. Detailed analysis of these knots and how they are ionizing the surrounding gas may shed more light on the nature and origins of \mygal. With access to images with lower seeing, velocity sliced H$\alpha$ maps of \mygal\ \citep[i.e., the analysis of Haro 11 in][]{Menacho2019} may help to probe the internal structure of the galaxy. We are interested in the sizes of these knots, since they are seeing limited in MUSE data. With good spatial resolution in the NIR, for example, we can search for sub-structures in these knots and model surface brightness profiles for them. Modelling the light profile of both galaxies is essential to improve our understanding on their detailed structure, which can trace mergers in the past history of both systems. It also allows us to measure the Cluster Formation Efficiency, defined as the fraction of total stellar mass formed in clusters per unit time in a given age interval divided by the SFR of the region where the clusters are detected. This quantity relates to the SFR surface density to quantify the intrinsic connection between massive star cluster formation processes and the mean properties of the host galaxies and compare to other galaxies, including dwarf irregulars and BCDGs \citep{adamo2011,adamo2020}. This posterior analysis may help us to understand better how this metal deficient galaxy is interacting with Mrk~1172 by improving the accuracy in its distance estimate, and thus shed some light about the formation and evolution of such systems. All these efforts are, however, beyond the scope of this paper, and are left for a future publication.

\section*{Acknowledgements}
We thank the referee for constructive comments and suggestions that have helped to improve the paper. This work was supported by Brazilian funding agencies CNPq and CAPES and by the \emph{Programa de Pós Graduação em Física} (PPGFis) at UFRGS.  A.L. acknowledges the hospitality in the visit to PUC Chile and ESO during Jan/Feb 2018. The authors also acknowledge the suggestions from Angela Adamo, Angela Krabbe, Laerte Sodré, Marina Trevisan, and Cristina Furlanetto. ACS acknowledges funding from CNPq and the Rio Grande do Sul Research Foundation (FAPERGS) through grants CNPq-403580/2016-1, CNPq-11153/2018-6, PqG/FAPERGS-17/2551-0001, FAPERGS/CAPES 19/2551-0000696-9 and L'Or\'eal UNESCO ABC \emph{Para Mulheres na Ci\^encia}. E.J. acknowledges support from FONDECYT Postdoctoral Fellowship Project Nº 3180557 and FONDECYT Iniciaci\'on 2020 Project No. 11200263. RR thanks CNPq, CAPES and FAPERGS for partial financial support to this project. RAR acknowledges financial support from CNPq (302280/2019-7 ) and FAPERGS (17/2551-0001144-9).
This study was based on observations collected at the European Organisation for Astronomical Research in the Southern Hemisphere under ESO programme 099.B-0411(A), PI: Johnston. This research has made use of the NASA/IPAC Extragalactic Database (NED), which is operated by the Jet Propulsion Laboratory, California Institute of Technology, under contract with NASA.

\section*{Data Availability}
The data used in this paper is available at ESO Science Archive Facility under the program-ID of 099.B-0411(A).




\bibliographystyle{mnras}
\bibliography{ref.bib} 

\begin{thebibliography}{}
\makeatletter
\relax
\def\mn@urlcharsother{\let\do\@makeother \do\$\do\&\do\#\do\^\do\_\do\%\do\~}
\def\mn@doi{\begingroup\mn@urlcharsother \@ifnextchar [ {\mn@doi@}
  {\mn@doi@[]}}
\def\mn@doi@[#1]#2{\def\@tempa{#1}\ifx\@tempa\@empty \href
  {http://dx.doi.org/#2} {doi:#2}\else \href {http://dx.doi.org/#2} {#1}\fi
  \endgroup}
\def\mn@eprint#1#2{\mn@eprint@#1:#2::\@nil}
\def\mn@eprint@arXiv#1{\href {http://arxiv.org/abs/#1} {{\tt arXiv:#1}}}
\def\mn@eprint@dblp#1{\href {http://dblp.uni-trier.de/rec/bibtex/#1.xml}
  {dblp:#1}}
\def\mn@eprint@#1:#2:#3:#4\@nil{\def\@tempa {#1}\def\@tempb {#2}\def\@tempc
  {#3}\ifx \@tempc \@empty \let \@tempc \@tempb \let \@tempb \@tempa \fi \ifx
  \@tempb \@empty \def\@tempb {arXiv}\fi \@ifundefined
  {mn@eprint@\@tempb}{\@tempb:\@tempc}{\expandafter \expandafter \csname
  mn@eprint@\@tempb\endcsname \expandafter{\@tempc}}}

\bibitem[\protect\citeauthoryear{{Adamo}, {{\"O}stlin}  \&
  {Zackrisson}}{{Adamo} et~al.}{2011}]{adamo2011}
{Adamo} A.,  {{\"O}stlin} G.,   {Zackrisson} E.,  2011, \mn@doi [\mnras]
  {10.1111/j.1365-2966.2011.19377.x}, \href
  {https://ui.adsabs.harvard.edu/abs/2011MNRAS.417.1904A} {417, 1904}

\bibitem[\protect\citeauthoryear{{Adamo} et~al.,}{{Adamo}
  et~al.}{2020}]{adamo2020}
{Adamo} A.,  et~al., 2020, \mn@doi [\ssr] {10.1007/s11214-020-00690-x}, \href
  {https://ui.adsabs.harvard.edu/abs/2020SSRv..216...69A} {216, 69}

\bibitem[\protect\citeauthoryear{{Ahn} et~al.,}{{Ahn} et~al.}{2012}]{SDSS2012}
{Ahn} C.~P.,  et~al., 2012, \mn@doi [\apjs] {10.1088/0067-0049/203/2/21}, \href
  {https://ui.adsabs.harvard.edu/abs/2012ApJS..203...21A} {203, 21}

\bibitem[\protect\citeauthoryear{{Allen}, {Groves}, {Dopita}, {Sutherland}  \&
  {Kewley}}{{Allen} et~al.}{2008}]{allen2008}
{Allen} M.~G.,  {Groves} B.~A.,  {Dopita} M.~A.,  {Sutherland} R.~S.,
  {Kewley} L.~J.,  2008, \mn@doi [\apjs] {10.1086/589652}, \href
  {https://ui.adsabs.harvard.edu/abs/2008ApJS..178...20A} {178, 20}

\bibitem[\protect\citeauthoryear{{Alloin}, {Collin-Souffrin}, {Joly}  \&
  {Vigroux}}{{Alloin} et~al.}{1979}]{alloin1979}
{Alloin} D.,  {Collin-Souffrin} S.,  {Joly} M.,   {Vigroux} L.,  1979, \aap,
  \href {https://ui.adsabs.harvard.edu/abs/1979A&A....78..200A} {78, 200}

\bibitem[\protect\citeauthoryear{{Aloisi} et~al.,}{{Aloisi}
  et~al.}{2007}]{aloisi2007}
{Aloisi} A.,  et~al., 2007, \mn@doi [\apjl] {10.1086/522368}, \href
  {https://ui.adsabs.harvard.edu/abs/2007ApJ...667L.151A} {667, L151}

\bibitem[\protect\citeauthoryear{{Andrews} \& {Martini}}{{Andrews} \&
  {Martini}}{2013}]{andrews2013}
{Andrews} B.~H.,  {Martini} P.,  2013, \mn@doi [\apj]
  {10.1088/0004-637X/765/2/140}, \href
  {https://ui.adsabs.harvard.edu/abs/2013ApJ...765..140A} {765, 140}

\bibitem[\protect\citeauthoryear{{Bacon} et~al.,}{{Bacon}
  et~al.}{2010}]{bacon2010}
{Bacon} R.,  et~al., 2010, in {McLean} I.~S.,  {Ramsay} S.~K.,   {Takami} H.,
  eds,  Society of Photo-Optical Instrumentation Engineers (SPIE) Conference
  Series Vol. 7735, Ground-based and Airborne Instrumentation for Astronomy
  III. p. 773508, \mn@doi{10.1117/12.856027}

\bibitem[\protect\citeauthoryear{{Baldwin}, {Phillips}  \&
  {Terlevich}}{{Baldwin} et~al.}{1981}]{baldwin1981}
{Baldwin} J.~A.,  {Phillips} M.~M.,   {Terlevich} R.,  1981, \mn@doi [\pasp]
  {10.1086/130766}, \href
  {https://ui.adsabs.harvard.edu/abs/1981PASP...93....5B} {93, 5}

\bibitem[\protect\citeauthoryear{{Bekki} \& {Stanimirovi{\'c}}}{{Bekki} \&
  {Stanimirovi{\'c}}}{2009}]{bekki2009}
{Bekki} K.,  {Stanimirovi{\'c}} S.,  2009, \mn@doi [\mnras]
  {10.1111/j.1365-2966.2009.14514.x}, \href
  {https://ui.adsabs.harvard.edu/abs/2009MNRAS.395..342B} {395, 342}

\bibitem[\protect\citeauthoryear{{Bernard} et~al.,}{{Bernard}
  et~al.}{2010}]{bernard2010}
{Bernard} E.~J.,  et~al., 2010, \mn@doi [\apj] {10.1088/0004-637X/712/2/1259},
  \href {https://ui.adsabs.harvard.edu/abs/2010ApJ...712.1259B} {712, 1259}

\bibitem[\protect\citeauthoryear{{Bertola}, {Bettoni}, {Danziger}, {Sadler},
  {Sparke}  \& {de Zeeuw}}{{Bertola} et~al.}{1991}]{bertola1991}
{Bertola} F.,  {Bettoni} D.,  {Danziger} J.,  {Sadler} E.,  {Sparke} L.,   {de
  Zeeuw} T.,  1991, \mn@doi [\apj] {10.1086/170058}, \href
  {https://ui.adsabs.harvard.edu/abs/1991ApJ...373..369B} {373, 369}

\bibitem[\protect\citeauthoryear{{Cardelli}, {Clayton}  \& {Mathis}}{{Cardelli}
  et~al.}{1989}]{cardelli1989}
{Cardelli} J.~A.,  {Clayton} G.~C.,   {Mathis} J.~S.,  1989, \mn@doi [\apj]
  {10.1086/167900}, \href
  {https://ui.adsabs.harvard.edu/abs/1989ApJ...345..245C} {345, 245}

\bibitem[\protect\citeauthoryear{{Ceverino}, {S{\'a}nchez Almeida}, {Mu{\~n}oz
  Tu{\~n}{\'o}n}, {Dekel}, {Elmegreen}, {Elmegreen}  \& {Primack}}{{Ceverino}
  et~al.}{2016}]{ceverino2016}
{Ceverino} D.,  {S{\'a}nchez Almeida} J.,  {Mu{\~n}oz Tu{\~n}{\'o}n} C.,
  {Dekel} A.,  {Elmegreen} B.~G.,  {Elmegreen} D.~M.,   {Primack} J.,  2016,
  \mn@doi [\mnras] {10.1093/mnras/stw064}, \href
  {https://ui.adsabs.harvard.edu/abs/2016MNRAS.457.2605C} {457, 2605}

\bibitem[\protect\citeauthoryear{{Cid Fernandes} \& {Gonz{\'a}lez
  Delgado}}{{Cid Fernandes} \& {Gonz{\'a}lez Delgado}}{2010}]{cid2010}
{Cid Fernandes} R.,  {Gonz{\'a}lez Delgado} R.~M.,  2010, \mn@doi [\mnras]
  {10.1111/j.1365-2966.2009.16153.x}, \href
  {https://ui.adsabs.harvard.edu/abs/2010MNRAS.403..780C} {403, 780}

\bibitem[\protect\citeauthoryear{{Cid Fernandes}, {Mateus}, {Sodr{\'e}},
  {Stasi{\'n}ska}  \& {Gomes}}{{Cid Fernandes} et~al.}{2005}]{cid2005}
{Cid Fernandes} R.,  {Mateus} A.,  {Sodr{\'e}} L.,  {Stasi{\'n}ska} G.,
  {Gomes} J.~M.,  2005, \mn@doi [\mnras] {10.1111/j.1365-2966.2005.08752.x},
  \href {https://ui.adsabs.harvard.edu/abs/2005MNRAS.358..363C} {358, 363}

\bibitem[\protect\citeauthoryear{{Cid Fernandes} et~al.,}{{Cid Fernandes}
  et~al.}{2013}]{cid2013}
{Cid Fernandes} R.,  et~al., 2013, \mn@doi [\aap]
  {10.1051/0004-6361/201220616}, \href
  {https://ui.adsabs.harvard.edu/abs/2013A&A...557A..86C} {557, A86}

\bibitem[\protect\citeauthoryear{{Cid Fernandes} et~al.,}{{Cid Fernandes}
  et~al.}{2014}]{cid2014}
{Cid Fernandes} R.,  et~al., 2014, \mn@doi [\aap]
  {10.1051/0004-6361/201321692}, \href
  {https://ui.adsabs.harvard.edu/abs/2014A&A...561A.130C} {561, A130}

\bibitem[\protect\citeauthoryear{{Croxall}, {van Zee}, {Lee}, {Skillman},
  {Lee}, {C{\^o}t{\'e}}, {Kennicutt}  \& {Miller}}{{Croxall}
  et~al.}{2009}]{croxall2009}
{Croxall} K.~V.,  {van Zee} L.,  {Lee} H.,  {Skillman} E.~D.,  {Lee} J.~C.,
  {C{\^o}t{\'e}} S.,  {Kennicutt} Robert~C. J.,   {Miller} B.~W.,  2009,
  \mn@doi [\apj] {10.1088/0004-637X/705/1/723}, \href
  {https://ui.adsabs.harvard.edu/abs/2009ApJ...705..723C} {705, 723}

\bibitem[\protect\citeauthoryear{{Dalcanton}}{{Dalcanton}}{2007}]{dalcanton2007}
{Dalcanton} J.~J.,  2007, \mn@doi [\apj] {10.1086/508913}, \href
  {https://ui.adsabs.harvard.edu/abs/2007ApJ...658..941D} {658, 941}

\bibitem[\protect\citeauthoryear{{Dekel}, {Sari}  \& {Ceverino}}{{Dekel}
  et~al.}{2009}]{dekel2009}
{Dekel} A.,  {Sari} R.,   {Ceverino} D.,  2009, \mn@doi [\apj]
  {10.1088/0004-637X/703/1/785}, \href
  {https://ui.adsabs.harvard.edu/abs/2009ApJ...703..785D} {703, 785}

\bibitem[\protect\citeauthoryear{{Delli Veneri}, {Cavuoti}, {Brescia}, {Longo}
  \& {Riccio}}{{Delli Veneri} et~al.}{2019}]{delli2019}
{Delli Veneri} M.,  {Cavuoti} S.,  {Brescia} M.,  {Longo} G.,   {Riccio} G.,
  2019, \mn@doi [\mnras] {10.1093/mnras/stz856}, \href
  {https://ui.adsabs.harvard.edu/abs/2019MNRAS.486.1377D} {486, 1377}

\bibitem[\protect\citeauthoryear{{Digby} et~al.,}{{Digby}
  et~al.}{2019}]{digby2019}
{Digby} R.,  et~al., 2019, \mn@doi [\mnras] {10.1093/mnras/stz745}, \href
  {https://ui.adsabs.harvard.edu/abs/2019MNRAS.485.5423D} {485, 5423}

\bibitem[\protect\citeauthoryear{{Dopita}, {Kewley}, {Sutherland}  \&
  {Nicholls}}{{Dopita} et~al.}{2016}]{dopita2016}
{Dopita} M.~A.,  {Kewley} L.~J.,  {Sutherland} R.~S.,   {Nicholls} D.~C.,
  2016, \mn@doi [\apss] {10.1007/s10509-016-2657-8}, \href
  {https://ui.adsabs.harvard.edu/abs/2016Ap&SS.361...61D} {361, 61}

\bibitem[\protect\citeauthoryear{{ESO CPL Development Team}}{{ESO CPL
  Development Team}}{2015}]{ESOteam2015}
{ESO CPL Development Team} 2015, {EsoRex: ESO Recipe Execution Tool}
  (\mn@eprint {ascl} {1504.003})

\bibitem[\protect\citeauthoryear{{Fattahi}, {Navarro}, {Frenk}, {Oman},
  {Sawala}  \& {Schaller}}{{Fattahi} et~al.}{2018}]{fattahi2018}
{Fattahi} A.,  {Navarro} J.~F.,  {Frenk} C.~S.,  {Oman} K.~A.,  {Sawala} T.,
  {Schaller} M.,  2018, \mn@doi [\mnras] {10.1093/mnras/sty408}, \href
  {https://ui.adsabs.harvard.edu/abs/2018MNRAS.476.3816F} {476, 3816}

\bibitem[\protect\citeauthoryear{{Gallart} et~al.,}{{Gallart}
  et~al.}{2015}]{gallart2015}
{Gallart} C.,  et~al., 2015, \mn@doi [\apjl] {10.1088/2041-8205/811/2/L18},
  \href {https://ui.adsabs.harvard.edu/abs/2015ApJ...811L..18G} {811, L18}

\bibitem[\protect\citeauthoryear{{Gibson} \& {Matteucci}}{{Gibson} \&
  {Matteucci}}{1997}]{gibson1997}
{Gibson} B.~K.,  {Matteucci} F.,  1997, \mn@doi [\apj] {10.1086/303513}, \href
  {https://ui.adsabs.harvard.edu/abs/1997ApJ...475...47G} {475, 47}

\bibitem[\protect\citeauthoryear{{Gil de Paz}, {Madore}  \& {Pevunova}}{{Gil de
  Paz} et~al.}{2003}]{gildepaz2003}
{Gil de Paz} A.,  {Madore} B.~F.,   {Pevunova} O.,  2003, \mn@doi [\apjs]
  {10.1086/374737}, \href
  {https://ui.adsabs.harvard.edu/abs/2003ApJS..147...29G} {147, 29}

\bibitem[\protect\citeauthoryear{{Gonz{\'a}lez Delgado} \& {Cid
  Fernandes}}{{Gonz{\'a}lez Delgado} \& {Cid Fernandes}}{2010}]{delgado2010}
{Gonz{\'a}lez Delgado} R.~M.,  {Cid Fernandes} R.,  2010, \mn@doi [\mnras]
  {10.1111/j.1365-2966.2009.16152.x}, \href
  {https://ui.adsabs.harvard.edu/abs/2010MNRAS.403..797G} {403, 797}

\bibitem[\protect\citeauthoryear{{Gonz{\'a}lez-Samaniego}, {Col{\'\i}n},
  {Avila-Reese}, {Rodr{\'\i}guez-Puebla}  \&
  {Valenzuela}}{{Gonz{\'a}lez-Samaniego} et~al.}{2014}]{gonzalez2014}
{Gonz{\'a}lez-Samaniego} A.,  {Col{\'\i}n} P.,  {Avila-Reese} V.,
  {Rodr{\'\i}guez-Puebla} A.,   {Valenzuela} O.,  2014, \mn@doi [\apj]
  {10.1088/0004-637X/785/1/58}, \href
  {https://ui.adsabs.harvard.edu/abs/2014ApJ...785...58G} {785, 58}

\bibitem[\protect\citeauthoryear{{Hinshaw} et~al.,}{{Hinshaw}
  et~al.}{2013}]{hinshaw2013}
{Hinshaw} G.,  et~al., 2013, \mn@doi [\apjs] {10.1088/0067-0049/208/2/19},
  \href {https://ui.adsabs.harvard.edu/abs/2013ApJS..208...19H} {208, 19}

\bibitem[\protect\citeauthoryear{{Jordi}, {Grebel}  \& {Ammon}}{{Jordi}
  et~al.}{2006}]{jordi2006}
{Jordi} K.,  {Grebel} E.~K.,   {Ammon} K.,  2006, \mn@doi [\aap]
  {10.1051/0004-6361:20066082}, \href
  {https://ui.adsabs.harvard.edu/abs/2006A&A...460..339J} {460, 339}

\bibitem[\protect\citeauthoryear{{Kauffmann} et~al.,}{{Kauffmann}
  et~al.}{2003}]{kauffmann2003}
{Kauffmann} G.,  et~al., 2003, \mn@doi [\mnras]
  {10.1111/j.1365-2966.2003.07154.x}, \href
  {https://ui.adsabs.harvard.edu/abs/2003MNRAS.346.1055K} {346, 1055}

\bibitem[\protect\citeauthoryear{{Kazantzidis}, {{\L}okas}, {Callegari},
  {Mayer}  \& {Moustakas}}{{Kazantzidis} et~al.}{2011}]{kazantzidis2010}
{Kazantzidis} S.,  {{\L}okas} E.~L.,  {Callegari} S.,  {Mayer} L.,
  {Moustakas} L.~A.,  2011, \mn@doi [\apj] {10.1088/0004-637X/726/2/98}, \href
  {https://ui.adsabs.harvard.edu/abs/2011ApJ...726...98K} {726, 98}

\bibitem[\protect\citeauthoryear{{Kennicutt}}{{Kennicutt}}{1998}]{kennicutt1998}
{Kennicutt} Robert~C. J.,  1998, \mn@doi [\araa]
  {10.1146/annurev.astro.36.1.189}, \href
  {https://ui.adsabs.harvard.edu/abs/1998ARA&A..36..189K} {36, 189}

\bibitem[\protect\citeauthoryear{{Kewley}, {Dopita}, {Sutherland}, {Heisler}
  \& {Trevena}}{{Kewley} et~al.}{2001}]{kewley2001}
{Kewley} L.~J.,  {Dopita} M.~A.,  {Sutherland} R.~S.,  {Heisler} C.~A.,
  {Trevena} J.,  2001, \mn@doi [\apj] {10.1086/321545}, \href
  {https://ui.adsabs.harvard.edu/abs/2001ApJ...556..121K} {556, 121}

\bibitem[\protect\citeauthoryear{{Kewley}, {Groves}, {Kauffmann}  \&
  {Heckman}}{{Kewley} et~al.}{2006}]{kewley2006}
{Kewley} L.~J.,  {Groves} B.,  {Kauffmann} G.,   {Heckman} T.,  2006, \mn@doi
  [\mnras] {10.1111/j.1365-2966.2006.10859.x}, \href
  {https://ui.adsabs.harvard.edu/abs/2006MNRAS.372..961K} {372, 961}

\bibitem[\protect\citeauthoryear{{Kewley}, {Zahid}, {Geller}, {Dopita}, {Hwang}
   \& {Fabricant}}{{Kewley} et~al.}{2015}]{Kewley2015}
{Kewley} L.~J.,  {Zahid} H.~J.,  {Geller} M.~J.,  {Dopita} M.~A.,  {Hwang}
  H.~S.,   {Fabricant} D.,  2015, \mn@doi [\apjl]
  {10.1088/2041-8205/812/2/L20}, \href
  {https://ui.adsabs.harvard.edu/abs/2015ApJ...812L..20K} {812, L20}

\bibitem[\protect\citeauthoryear{{Kirby}, {Cohen}, {Guhathakurta}, {Cheng},
  {Bullock}  \& {Gallazzi}}{{Kirby} et~al.}{2013}]{kirby2013}
{Kirby} E.~N.,  {Cohen} J.~G.,  {Guhathakurta} P.,  {Cheng} L.,  {Bullock}
  J.~S.,   {Gallazzi} A.,  2013, \mn@doi [\apj] {10.1088/0004-637X/779/2/102},
  \href {https://ui.adsabs.harvard.edu/abs/2013ApJ...779..102K} {779, 102}

\bibitem[\protect\citeauthoryear{{Kormendy} \& {Bender}}{{Kormendy} \&
  {Bender}}{2012}]{kormendy2011}
{Kormendy} J.,  {Bender} R.,  2012, \mn@doi [\apjs]
  {10.1088/0067-0049/198/1/2}, \href
  {https://ui.adsabs.harvard.edu/abs/2012ApJS..198....2K} {198, 2}

\bibitem[\protect\citeauthoryear{{Kunth} \& {{\"O}stlin}}{{Kunth} \&
  {{\"O}stlin}}{2000}]{kunth2000}
{Kunth} D.,  {{\"O}stlin} G.,  2000, \mn@doi [\aapr] {10.1007/s001590000005},
  \href {https://ui.adsabs.harvard.edu/abs/2000A&ARv..10....1K} {10, 1}

\bibitem[\protect\citeauthoryear{{Lequeux}, {Peimbert}, {Rayo}, {Serrano}  \&
  {Torres-Peimbert}}{{Lequeux} et~al.}{1979}]{lequeux1979}
{Lequeux} J.,  {Peimbert} M.,  {Rayo} J.~F.,  {Serrano} A.,   {Torres-Peimbert}
  S.,  1979, \aap, \href
  {https://ui.adsabs.harvard.edu/abs/1979A&A....80..155L} {500, 145}

\bibitem[\protect\citeauthoryear{{Mallmann} et~al.,}{{Mallmann}
  et~al.}{2018}]{nicolas2018}
{Mallmann} N.~D.,  et~al., 2018, \mn@doi [\mnras] {10.1093/mnras/sty1364},
  \href {https://ui.adsabs.harvard.edu/abs/2018MNRAS.478.5491M} {478, 5491}

\bibitem[\protect\citeauthoryear{{Marino} et~al.,}{{Marino}
  et~al.}{2013}]{marino2013}
{Marino} R.~A.,  et~al., 2013, \mn@doi [\aap] {10.1051/0004-6361/201321956},
  \href {https://ui.adsabs.harvard.edu/abs/2013A&A...559A.114M} {559, A114}

\bibitem[\protect\citeauthoryear{{Mashchenko}, {Wadsley}  \&
  {Couchman}}{{Mashchenko} et~al.}{2008}]{mashchenko2008}
{Mashchenko} S.,  {Wadsley} J.,   {Couchman} H.~M.~P.,  2008, \mn@doi [Science]
  {10.1126/science.1148666}, \href
  {https://ui.adsabs.harvard.edu/abs/2008Sci...319..174M} {319, 174}

\bibitem[\protect\citeauthoryear{{Mateo}}{{Mateo}}{1998}]{mateo1998}
{Mateo} M.~L.,  1998, \mn@doi [\araa] {10.1146/annurev.astro.36.1.435}, \href
  {https://ui.adsabs.harvard.edu/abs/1998ARA&A..36..435M} {36, 435}

\bibitem[\protect\citeauthoryear{{Mayer}, {Governato}, {Colpi}, {Moore},
  {Quinn}, {Wadsley}, {Stadel}  \& {Lake}}{{Mayer} et~al.}{2001}]{mayer2001}
{Mayer} L.,  {Governato} F.,  {Colpi} M.,  {Moore} B.,  {Quinn} T.,  {Wadsley}
  J.,  {Stadel} J.,   {Lake} G.,  2001, \mn@doi [\apjl] {10.1086/318898}, \href
  {https://ui.adsabs.harvard.edu/abs/2001ApJ...547L.123M} {547, L123}

\bibitem[\protect\citeauthoryear{{Menacho} et~al.,}{{Menacho}
  et~al.}{2019}]{Menacho2019}
{Menacho} V.,  et~al., 2019, \mn@doi [\mnras] {10.1093/mnras/stz1414}, \href
  {https://ui.adsabs.harvard.edu/abs/2019MNRAS.487.3183M} {487, 3183}

\bibitem[\protect\citeauthoryear{{Monreal-Ibero}, {Rela{\~n}o}, {Kehrig},
  {P{\'e}rez-Montero}, {V{\'\i}lchez}, {Kelz}, {Roth}  \&
  {Streicher}}{{Monreal-Ibero} et~al.}{2011}]{Monreal_Ibero2011}
{Monreal-Ibero} A.,  {Rela{\~n}o} M.,  {Kehrig} C.,  {P{\'e}rez-Montero} E.,
  {V{\'\i}lchez} J.~M.,  {Kelz} A.,  {Roth} M.~M.,   {Streicher} O.,  2011,
  \mn@doi [\mnras] {10.1111/j.1365-2966.2011.18300.x}, \href
  {https://ui.adsabs.harvard.edu/abs/2011MNRAS.413.2242M} {413, 2242}

\bibitem[\protect\citeauthoryear{{Nagao}, {Maiolino}  \& {Marconi}}{{Nagao}
  et~al.}{2006}]{nagao2006}
{Nagao} T.,  {Maiolino} R.,   {Marconi} A.,  2006, \mn@doi [\aap]
  {10.1051/0004-6361:20065216}, \href
  {https://ui.adsabs.harvard.edu/abs/2006A&A...459...85N} {459, 85}

\bibitem[\protect\citeauthoryear{{Navarro}, {Eke}  \& {Frenk}}{{Navarro}
  et~al.}{1996}]{navarro1996}
{Navarro} J.~F.,  {Eke} V.~R.,   {Frenk} C.~S.,  1996, \mn@doi [\mnras]
  {10.1093/mnras/283.3.L72}, \href
  {https://ui.adsabs.harvard.edu/abs/1996MNRAS.283L..72N} {283, L72}

\bibitem[\protect\citeauthoryear{{O'Donnell}}{{O'Donnell}}{1994}]{odonnell1994}
{O'Donnell} J.~E.,  1994, \mn@doi [\apj] {10.1086/173713}, \href
  {https://ui.adsabs.harvard.edu/abs/1994ApJ...422..158O} {422, 158}

\bibitem[\protect\citeauthoryear{{Omand}, {Balogh}  \& {Poggianti}}{{Omand}
  et~al.}{2014}]{omand2014}
{Omand} C. M.~B.,  {Balogh} M.~L.,   {Poggianti} B.~M.,  2014, \mn@doi [\mnras]
  {10.1093/mnras/stu331}, \href
  {https://ui.adsabs.harvard.edu/abs/2014MNRAS.440..843O} {440, 843}

\bibitem[\protect\citeauthoryear{Osterbrock \& Ferland}{Osterbrock \&
  Ferland}{2006}]{osterbrock2006}
Osterbrock D.~E.,  Ferland G.~J.,  2006, Astrophysics Of Gas Nebulae and Active
  Galactic Nuclei.
University science books

\bibitem[\protect\citeauthoryear{{P{\'e}rez-Montero} \&
  {Contini}}{{P{\'e}rez-Montero} \& {Contini}}{2009}]{perez2009}
{P{\'e}rez-Montero} E.,  {Contini} T.,  2009, \mn@doi [\mnras]
  {10.1111/j.1365-2966.2009.15145.x}, \href
  {https://ui.adsabs.harvard.edu/abs/2009MNRAS.398..949P} {398, 949}

\bibitem[\protect\citeauthoryear{{Pettini} \& {Pagel}}{{Pettini} \&
  {Pagel}}{2004}]{pettini2004}
{Pettini} M.,  {Pagel} B. E.~J.,  2004, \mn@doi [\mnras]
  {10.1111/j.1365-2966.2004.07591.x}, \href
  {https://ui.adsabs.harvard.edu/abs/2004MNRAS.348L..59P} {348, L59}

\bibitem[\protect\citeauthoryear{{Pilyugin} \& {Ferrini}}{{Pilyugin} \&
  {Ferrini}}{2000}]{pilyugin2000}
{Pilyugin} L.~S.,  {Ferrini} F.,  2000, \mn@doi [\nar]
  {10.1016/S1387-6473(00)00057-9}, \href
  {https://ui.adsabs.harvard.edu/abs/2000NewAR..44..335P} {44, 335}

\bibitem[\protect\citeauthoryear{{Reines}, {Sivakoff}, {Johnson}  \&
  {Brogan}}{{Reines} et~al.}{2011}]{reines2011}
{Reines} A.~E.,  {Sivakoff} G.~R.,  {Johnson} K.~E.,   {Brogan} C.~L.,  2011,
  \mn@doi [\nat] {10.1038/nature09724}, \href
  {https://ui.adsabs.harvard.edu/abs/2011Natur.470...66R} {470, 66}

\bibitem[\protect\citeauthoryear{{Revaz} \& {Jablonka}}{{Revaz} \&
  {Jablonka}}{2018}]{revaz2018}
{Revaz} Y.,  {Jablonka} P.,  2018, \mn@doi [\aap]
  {10.1051/0004-6361/201832669}, \href
  {https://ui.adsabs.harvard.edu/abs/2018A&A...616A..96R} {616, A96}

\bibitem[\protect\citeauthoryear{{Riffel}, {Pastoriza}, {Rodr{\'\i}guez-Ardila}
   \& {Bonatto}}{{Riffel} et~al.}{2009}]{riffel2009}
{Riffel} R.,  {Pastoriza} M.~G.,  {Rodr{\'\i}guez-Ardila} A.,   {Bonatto} C.,
  2009, \mn@doi [\mnras] {10.1111/j.1365-2966.2009.15448.x}, \href
  {https://ui.adsabs.harvard.edu/abs/2009MNRAS.400..273R} {400, 273}

\bibitem[\protect\citeauthoryear{{Riffel} et~al.,}{{Riffel}
  et~al.}{2021}]{riffel2021}
{Riffel} R.,  et~al., 2021, \mn@doi [\mnras] {10.1093/mnras/staa3907}, \href
  {https://ui.adsabs.harvard.edu/abs/2021MNRAS.501.4064R} {501, 4064}

\bibitem[\protect\citeauthoryear{{Rubin}, {Ford}  \& {Whitmore}}{{Rubin}
  et~al.}{1984}]{rubin1984}
{Rubin} V.~C.,  {Ford} W.~K. J.,   {Whitmore} B.~C.,  1984, \mn@doi [\apjl]
  {10.1086/184276}, \href
  {https://ui.adsabs.harvard.edu/abs/1984ApJ...281L..21R} {281, L21}

\bibitem[\protect\citeauthoryear{Ruschel-Dutra \& de Oliveira}{Ruschel-Dutra \&
  de~Oliveira}{2020}]{ifscube}
Ruschel-Dutra D.,  de Oliveira B.~D.,  2020, danielrd6/ifscube: Modeling,
  \mn@doi{10.5281/zenodo.4065550}, \url
  {https://doi.org/10.5281/zenodo.4065550}

\bibitem[\protect\citeauthoryear{Ryden \& Pogge}{Ryden \&
  Pogge}{2015}]{ryden2015}
Ryden B.,  Pogge R.,  2015, Interstellar and Intergalactic Medium.
The Ohio State University

\bibitem[\protect\citeauthoryear{{S{\'a}nchez Almeida} et~al.,}{{S{\'a}nchez
  Almeida} et~al.}{2015}]{sanchezalmeida2015}
{S{\'a}nchez Almeida} J.,  et~al., 2015, \mn@doi [\apjl]
  {10.1088/2041-8205/810/2/L15}, \href
  {https://ui.adsabs.harvard.edu/abs/2015ApJ...810L..15S} {810, L15}

\bibitem[\protect\citeauthoryear{{Schlegel}, {Finkbeiner}  \&
  {Davis}}{{Schlegel} et~al.}{1998}]{schlegel1998}
{Schlegel} D.~J.,  {Finkbeiner} D.~P.,   {Davis} M.,  1998, \mn@doi [\apj]
  {10.1086/305772}, \href
  {https://ui.adsabs.harvard.edu/abs/1998ApJ...500..525S} {500, 525}

\bibitem[\protect\citeauthoryear{{Skillman} \& {Bender}}{{Skillman} \&
  {Bender}}{1995}]{skillman1995}
{Skillman} E.~D.,  {Bender} R.,  1995, in {Pena} M.,  {Kurtz} S.,  eds,
  Revista Mexicana de Astronomia y Astrofisica Conference Series Vol. 3,
  Revista Mexicana de Astronomia y Astrofisica Conference Series. p.~25

\bibitem[\protect\citeauthoryear{{Skillman}, {Kennicutt}  \&
  {Hodge}}{{Skillman} et~al.}{1989}]{skillman1989}
{Skillman} E.~D.,  {Kennicutt} R.~C.,   {Hodge} P.~W.,  1989, \mn@doi [\apj]
  {10.1086/168178}, \href
  {https://ui.adsabs.harvard.edu/abs/1989ApJ...347..875S} {347, 875}

\bibitem[\protect\citeauthoryear{{Soto}, {Lilly}, {Bacon}, {Richard}  \&
  {Conseil}}{{Soto} et~al.}{2016}]{zap2016}
{Soto} K.~T.,  {Lilly} S.~J.,  {Bacon} R.,  {Richard} J.,   {Conseil} S.,
  2016, \mn@doi [\mnras] {10.1093/mnras/stw474}, \href
  {https://ui.adsabs.harvard.edu/abs/2016MNRAS.458.3210S} {458, 3210}

\bibitem[\protect\citeauthoryear{{Steffen}, {Prakapavi{\v{c}}ius}, {Caffau},
  {Ludwig}, {Bonifacio}, {Cayrel}, {Ku{\v{c}}inskas}  \&
  {Livingston}}{{Steffen} et~al.}{2015}]{steffen2015}
{Steffen} M.,  {Prakapavi{\v{c}}ius} D.,  {Caffau} E.,  {Ludwig} H.~G.,
  {Bonifacio} P.,  {Cayrel} R.,  {Ku{\v{c}}inskas} A.,   {Livingston} W.~C.,
  2015, \mn@doi [\aap] {10.1051/0004-6361/201526406}, \href
  {https://ui.adsabs.harvard.edu/abs/2015A&A...583A..57S} {583, A57}

\bibitem[\protect\citeauthoryear{{Steyrleithner}, {Hensler}  \&
  {Boselli}}{{Steyrleithner} et~al.}{2020}]{Steyrleithner20}
{Steyrleithner} P.,  {Hensler} G.,   {Boselli} A.,  2020, \mn@doi [\mnras]
  {10.1093/mnras/staa775}, \href
  {https://ui.adsabs.harvard.edu/abs/2020MNRAS.494.1114S} {494, 1114}

\bibitem[\protect\citeauthoryear{{Storchi-Bergmann}, {Calzetti}  \&
  {Kinney}}{{Storchi-Bergmann} et~al.}{1994}]{thaisa1994}
{Storchi-Bergmann} T.,  {Calzetti} D.,   {Kinney} A.~L.,  1994, \mn@doi [\apj]
  {10.1086/174345}, \href
  {https://ui.adsabs.harvard.edu/abs/1994ApJ...429..572S} {429, 572}

\bibitem[\protect\citeauthoryear{{Tremonti} et~al.,}{{Tremonti}
  et~al.}{2004}]{tremonti2004}
{Tremonti} C.~A.,  et~al., 2004, \mn@doi [\apj] {10.1086/423264}, \href
  {https://ui.adsabs.harvard.edu/abs/2004ApJ...613..898T} {613, 898}

\bibitem[\protect\citeauthoryear{{Vazdekis}, {S{\'a}nchez-Bl{\'a}zquez},
  {Falc{\'o}n-Barroso}, {Cenarro}, {Beasley}, {Cardiel}, {Gorgas}  \&
  {Peletier}}{{Vazdekis} et~al.}{2010}]{vazdekis2010}
{Vazdekis} A.,  {S{\'a}nchez-Bl{\'a}zquez} P.,  {Falc{\'o}n-Barroso} J.,
  {Cenarro} A.~J.,  {Beasley} M.~A.,  {Cardiel} N.,  {Gorgas} J.,   {Peletier}
  R.~F.,  2010, \mn@doi [\mnras] {10.1111/j.1365-2966.2010.16407.x}, \href
  {https://ui.adsabs.harvard.edu/abs/2010MNRAS.404.1639V} {404, 1639}

\bibitem[\protect\citeauthoryear{{Vazdekis}, {Koleva}, {Ricciardelli},
  {R{\"o}ck}  \& {Falc{\'o}n-Barroso}}{{Vazdekis} et~al.}{2016}]{vazdekis16}
{Vazdekis} A.,  {Koleva} M.,  {Ricciardelli} E.,  {R{\"o}ck} B.,
  {Falc{\'o}n-Barroso} J.,  2016, \mn@doi [\mnras] {10.1093/mnras/stw2231},
  \href {https://ui.adsabs.harvard.edu/abs/2016MNRAS.463.3409V} {463, 3409}

\bibitem[\protect\citeauthoryear{{Weilbacher} et~al.,}{{Weilbacher}
  et~al.}{2020}]{weilbacher2020}
{Weilbacher} P.~M.,  et~al., 2020, \mn@doi [\aap]
  {10.1051/0004-6361/202037855}, \href
  {https://ui.adsabs.harvard.edu/abs/2020A&A...641A..28W} {641, A28}

\bibitem[\protect\citeauthoryear{{Wylezalek} et~al.,}{{Wylezalek}
  et~al.}{2017}]{wylezalek2017}
{Wylezalek} D.,  et~al., 2017, \mn@doi [\mnras] {10.1093/mnras/stx246}, \href
  {https://ui.adsabs.harvard.edu/abs/2017MNRAS.467.2612W} {467, 2612}

\bibitem[\protect\citeauthoryear{{do Nascimento} et~al.,}{{do Nascimento}
  et~al.}{2019}]{nascimento2019}
{do Nascimento} J.~C.,  et~al., 2019, \mn@doi [\mnras] {10.1093/mnras/stz1083},
  \href {https://ui.adsabs.harvard.edu/abs/2019MNRAS.486.5075D} {486, 5075}

\bibitem[\protect\citeauthoryear{{van Zee}}{{van Zee}}{2001}]{vanzee2001}
{van Zee} L.,  2001, \mn@doi [\aj] {10.1086/319947}, \href
  {https://ui.adsabs.harvard.edu/abs/2001AJ....121.2003V} {121, 2003}

\bibitem[\protect\citeauthoryear{{van Zee} \& {Haynes}}{{van Zee} \&
  {Haynes}}{2006}]{vanzee2006}
{van Zee} L.,  {Haynes} M.~P.,  2006, \mn@doi [\apj] {10.1086/498017}, \href
  {https://ui.adsabs.harvard.edu/abs/2006ApJ...636..214V} {636, 214}

\bibitem[\protect\citeauthoryear{{van der Kruit} \& {Allen}}{{van der Kruit} \&
  {Allen}}{1978}]{kruit1978}
{van der Kruit} P.~C.,  {Allen} R.~J.,  1978, \mn@doi [\araa]
  {10.1146/annurev.aa.16.090178.000535}, \href
  {https://ui.adsabs.harvard.edu/abs/1978ARA&A..16..103V} {16, 103}

\makeatother
\end{thebibliography}




\appendix

\bsp	
\label{lastpage}
\end{document}